\documentclass[arps,prd,onecolumn,twoside,floatfix,noshowpacs,nofootinbib,hyperref]{revtex4}
\pdfoutput=1

\usepackage{color} 
\usepackage{amsmath}
\usepackage{comment}
\usepackage{bm}

\newcommand\beq{\begin{equation}}
\newcommand\eeq{\end{equation}}
\newcommand{\beqa}{\begin{eqnarray}}
\newcommand{\eeqa}{\end{eqnarray}}
\newcommand\beqn{\begin{eqnarray}}
\newcommand\eeqn{\end{eqnarray}}

\newcommand{\y}{$y$~}

\newcommand{\ba}{\begin{eqnarray}}
\newcommand{\ea}{\end{eqnarray}}
\newcommand{\be}{\begin{equation}}
\newcommand{\ee}{\end{equation}}

\newcommand\lsim{\mathrel{\rlap{\lower4pt\hbox{\hskip1pt$\sim$}}
        \raise1pt\hbox{$<$}}}
\newcommand\gsim{\mathrel{\rlap{\lower4pt\hbox{\hskip1pt$\sim$}}
        \raise1pt\hbox{$>$}}}

\newcommand{\jcap}{{J.~Cosm.~Astrop.~Phys.}}

\newcommand{\aap}{{Astron.~Astrophys.}}
\newcommand{\apjl}{{Astrophys.~J.~Lett.}}
\newcommand{\apjs}{{Astrophys.~J.~Supp.}}

\newcommand{\mnras}{{Mon.~Not.~R.~Astron.~Soc.}}
\usepackage{graphicx}
\usepackage[large]{subfigure}
\usepackage{amssymb, amsmath}
\usepackage[amssymb]{SIunits}
\usepackage{epstopdf}
\usepackage{hyperref}
\usepackage{url}
\usepackage{aas_macros}
\usepackage{natbib}

\defcitealias{PlanckLBG2013}{P13}

\begin{document}
\title{The Two-Halo Term in Stacked Thermal Sunyaev-Zel'dovich Measurements: Implications for Self-Similarity}
\author{J.~Colin Hill,\footnote{jch@ias.edu}$^{1}$ Eric J.~Baxter,$^{2}$ Adam Lidz,$^{2}$ Johnny P.~Greco,$^{3}$ and Bhuvnesh Jain$^{2}$}
\affiliation{$^{1}$Dept.~of Astronomy, Columbia University, New York, NY USA 10027 \\
$^{2}$Dept.~of Physics and Astronomy, University of Pennsylvania, Philadelphia, PA USA 19104 \\
$^{3}$Dept.~of Astrophysical Sciences, Princeton University, Princeton, NJ USA 08544}

\begin{abstract}
The relation between the mass and integrated electron pressure of
galaxy group and cluster halos can be probed by stacking maps of the
thermal Sunyaev-Zel'dovich (tSZ) effect. Perhaps surprisingly, recent
observational results have indicated that the scaling relation between
integrated pressure and mass follows the prediction of simple,
self-similar models down to halo masses as low as $10^{12.5} \,
M_{\odot}$. Hydrodynamical simulations that incorporate energetic
feedback processes suggest that gas should be depleted from such
low-mass halos, thus decreasing their tSZ signal relative to
self-similar predictions.  Here, we build on the modeling of Vikram, Lidz, and Jain~(2017) 
to evaluate the bias in the interpretation of stacked tSZ measurements due to the signal 
from correlated halos (the ``two-halo'' term), which has generally been
neglected in the literature. We fit theoretical models to a
measurement of the tSZ -- galaxy group cross-correlation function,
accounting explicitly for the one- and two- halo contributions. We
find moderate evidence of a deviation from self-similarity in the
pressure -- mass relation, even after marginalizing over conservative
miscentering effects. We explore pressure -- mass models with a break
at $10^{14} \, M_{\odot}$, as well as other variants. We discuss and
test for sources of uncertainty in our analysis, in particular a
possible bias in the halo mass estimates and the coarse resolution of
the Planck beam.  We compare our findings with earlier analyses by
exploring the extent to which halo isolation criteria can reduce the
two-halo contribution.  Finally, we show that ongoing third-generation
CMB experiments will explicitly resolve the one-halo term in low-mass
groups; our methodology can be applied to these upcoming data sets to
obtain a clear answer to the question of self-similarity and an
improved understanding of hot gas in low-mass halos.
\end{abstract}

\maketitle

\section{Introduction}
\label{sec:intro}

Simple models of cosmic structure formation based on gravitation alone predict nearly self-similar relations between halo mass and various thermodynamic quantities characterizing the gas in a halo potential, including the thermal gas pressure profile~(e.g.,~\cite{Kaiser1986,Komatsu-Seljak2001}).  A variety of observations at galaxy cluster mass scales, where gravity dominates the overall halo energy budget, have lent support to this picture, particularly at large cluster-centric radii~(e.g.,~\cite{Arnaud2010,Planck2013profile,Mantz2016}).  However, there are many physical processes that could cause deviations from self-similarity, particularly at lower mass scales, where energetic feedback from supernovae and active galactic nuclei (AGN), as well as other non-gravitational processes (e.g., turbulent pressure support, cosmic rays, or magnetic fields), significantly influence the thermodynamic state of the halo gas~(e.g.,~\cite{Borgani2004,Parrish2012,Battaglia2012YM,McCourt2013,LeBrun2014,Nelson2014}).  Indeed, X-ray observations have long indicated that groups and low-mass clusters do not precisely follow self-similar relations between, e.g., the X-ray luminosity and temperature~(e.g.,~\cite{Kaiser1991,Voit2005,Osmond2004,Sun2009}), although the X-ray data probe the innermost group/cluster regions, where non-gravitational effects are most important.  In the current picture, these non-self-similar observations are primarily explained by AGN feedback, which drives gas out of the inner cluster regions, thus reducing the X-ray luminosity and gas fraction~(e.g.,~\cite{Puchwein2008,Battaglia2013gasfrac}).  However, the distribution and thermodynamic properties of the gas at large halo-centric radii, as well as at sub-cluster mass scales, remain open questions.

The thermal Sunyaev-ZelÕdovich (tSZ) effect is a powerful probe of the thermodynamic state of the gas in and around halos.  The tSZ effect is the inverse-Compton scattering of cosmic microwave background (CMB) photons off hot electrons, which leads to a characteristic distortion in the spectrum of the CMB: at frequencies below (above) $\approx 217$ GHz, a decrement (increment) is observed in the CMB temperature at the location of the scattering electrons~\cite{Sunyaev-Zeldovich1970}.  The amplitude of the CMB temperature shift is characterized by the Compton-$y$ parameter, which is given by a line-of-sight integral of the electron pressure.  Measurements of the relation between the Compton-$y$ signal and halo mass, redshift, galaxy type, and other properties can thus shed light on the questions related to structure formation described above.  In the simple, self-similar halo model, the volume-integrated Compton-$y$ parameter $Y$ can be shown to scale with the halo mass $M$ as $Y \propto M^{5/3}$.  At galaxy cluster mass scales, numerical simulations and observations have generally yielded results fairly close to this prediction~(e.g.,~\cite{Arnaud2010,Battaglia2010,Battaglia2012,LeBrun2017,McCarthy2017,Kay2012,Planelles2017}).  At lower mass scales (e.g., $M \lesssim 10^{14} \, M_{\odot}$), this simple prediction is expected to fail due to the importance of non-gravitational processes.  This departure from self-similarity can be quantified by a power-law $Y$--$M$ relation with a slope deviating from~5/3, or, more realistically, by a broken power-law (or other shape) in which the $Y$ signal of halos below some critical ``break'' mass is suppressed relative to the self-similar expectation, while that at higher masses remains consistent with self-similarity (or, more generally, a power-law near~5/3).  We consider both the pure power-law and broken power-law possibilities in what follows.  Some theoretical models predict noticeable deviations from self-similarity in the $Y$--$M$ relation below a mass scale $\approx 10^{14} \, M_{\odot}$~(e.g.,~\cite{LeBrun2015,LeBrun2017,vandeVoort2016}), which we adopt as a canonical ``break mass'' in this paper.

In recent years, the combination of exquisite CMB data and large galaxy surveys has led to several studies aiming to use the tSZ signal of various halo samples to constrain the $Y$--$M$ relation, including stacking analyses~(e.g.,~\cite{Hand2011,PlanckMaxBCG2011,PlanckLBG2013,Greco2015,Jimeno2017}) and cross-correlation measurements~(e.g.,~\cite{Hill-Spergel2014,vWHM2014,Hojjati2016,Vikram2017}).  A noteworthy breakthrough was achieved by~\citet{PlanckLBG2013} (hereafter
\citetalias{PlanckLBG2013}), who were able to detect the tSZ signal of halos at unprecedentedly low mass scales ($M \approx 10^{12.5} \, M_{\odot}$) by stacking Planck data on a sample of ``locally brightest galaxies'' (LBGs) extracted from Sloan Digital Sky Survey (SDSS) data.  Perhaps surprisingly, their results indicated that self-similar predictions were consistent with the inferred $Y$--$M$ relation down to these low masses.  A subsequent analysis by~\citet{Greco2015} arrived at essentially the same conclusions (while using a slightly different methodology), although they found that some of the tSZ signal for the lowest-mass halos may have in fact been due to dust contamination.  However, both~\citet{Greco2015} and~\citet{LeBrun2015} pointed out that the analysis in~\citetalias{PlanckLBG2013} did not in fact measure the tSZ signal within $r_{500}$ (the radius enclosing a mass within which the mean density is 500 times the critical density), as originally claimed, but rather within a much larger aperture of radius $5r_{500}$.  The measurements within the larger aperture were rescaled to the smaller aperture by assuming the validity of the~\citet{Arnaud2010} pressure profile, despite the fact that this profile was only measured/calibrated for massive galaxy clusters.  \citet{LeBrun2015} explicitly showed that this rescaling could introduce significant biases in the results.  They argued that the~\citetalias{PlanckLBG2013} measurements were, in fact, inconsistent with a self-similar mass dependence, if a realistic pressure profile based on simulations with AGN feedback was adopted for the rescaling.  Further developments were presented in~\citet{Anderson2015}, who stacked ROSAT All-Sky Survey X-ray data on the LBG sample, finding a non-self-similar relation between the X-ray luminosity and halo mass, as well as~\citet{Wang2016}, who recalibrated the scaling relations for the LBG sample using weak lensing data to obtain halo masses.  In particular, the latter study found that the weak lensing calibration of the LBG sample still yielded a $Y$--$M$ relation consistent with the self-similar prediction, albeit subject to the large-aperture caveat described above for the $Y$ measurements.

In this work, we point out a missing component in the modeling of nearly all previous tSZ stacking studies, including those mentioned above: an explicit accounting for the two-halo term, i.e., the correlated tSZ signal due to objects other than the halo of interest.\footnote{Note that the two-halo term was explicitly modeled and measured in all tSZ--lensing cross-correlations presented to date~\cite{Hill-Spergel2014,vWHM2014,Hojjati2016}.  Here, we are explicitly focused on tSZ stacking or cross-correlation analyses on halo samples.}  \citet{Vikram2017} first explicitly measured the two-halo term in the tSZ--galaxy group cross-correlation function, and pointed out that it could dominate the total tSZ signal around low-mass halos, due to the strong mass dependence of the Compton-$y$ signal.  \citetalias{PlanckLBG2013} attempted to mitigate possible two-halo contributions by applying ``isolation criteria'' to the LBG sample extracted from SDSS.  We revisit these criteria in further detail below.  Nevertheless, unless the isolation criteria were near-perfect, some residual two-halo contribution is to be expected, which should be modeled in the analysis --- if not, a bias in the inferred $Y$--$M$ relation or pressure profile behavior will result.  Indeed, recently~\citet{Jimeno2017} presented a tSZ stacking analysis on a sample of SDSS redMaPPer galaxy clusters, without modeling the two-halo term.  Although the two-halo term likely does not dominate the signal for any objects in their sample, it could be responsible for their claimed evidence for a flatter pressure profile slope in the outer cluster regions, compared to that expected from theoretical predictions (see, e.g., Fig.~\ref{fig:data} below).

Similar analyses have also been presented for samples of quasars, with the primary goal of using the tSZ effect to detect or constrain any additional energy input due to AGN feedback in these systems (beyond the gravitational energy)~\cite{Ruan2015,Crichton2016,Verdier2016}.\footnote{Similar analyses have also been performed for massive elliptical galaxies (rather than quasars)~\cite{Spacek2016,Spacek2017}.}  Claims of evidence for feedback have been presented, but none of these studies have modeled the two-halo term, which could thus be responsible for (at least part of) the excess signal.  Indeed,~\citet{Cen-Safarzadeh2015} explicitly showed that the excess feedback energy claimed in the quasar tSZ stacking results of~\citet{Ruan2015} could be explained by two-halo contributions, by using the Millennium Simulation to construct tSZ maps according to various quasar halo occupation distributions.

Here, we present a method to analytically model the two-halo (and one-halo) tSZ signal in terms of the $y$-galaxy group cross-correlation function, following~\citet{Vikram2017}.  Using this methodology to fit various pressure profile models to $y$-group correlation function measurements, we find moderate evidence for a departure from self-similarity in the $Y$--$M$ relation, suggesting that the earlier results may have been biased by their neglect of the two-halo term.  We point out that the inferred $Y$--$M$ behavior can be sensitive to the parameterization adopted in the theoretical model, with simple power-law fits possibly obscuring evidence of non-self-similar behavior at low masses.  We investigate these results further by measuring the $y$-LBG cross-correlation function, while varying the LBG isolation criteria over wide ranges.  We find that the LBG signal in low-mass halos is not extremely robust to such variations.  We argue that future measurements of the tSZ signal from any halo samples should model and account for the two-halo term.

The remainder of this paper is organized as follows.  In \S\ref{sec:data}, we discuss the data sets used in this work and describe our cross-correlation measurements.  In \S\ref{sec:halo}, we review the theoretical halo models used to interpret the measurements.  \S\ref{sec:analysis} presents the results of fitting these models to the data, as well as an investigation of the role of isolation criteria in removing the two-halo term from stacked tSZ measurements.  We discuss the results and conclude with an outlook for upcoming measurements in \S\ref{sec:interp}.  Throughout, we assume the following cosmological parameters in our theoretical calculations: 
$\Omega_m=0.27$, $\Omega_\Lambda=0.73$, $\Omega_b=0.044$, $h=0.7$, $\sigma_8 = 0.8$, $n_s=1$.

\section{Data and Measurements}
\label{sec:data}

Our analysis makes use of data from both the Planck satellite and the
Sloan Digital Sky Survey (SDSS).  We measure the $y$-group
cross-correlation using the Compton-$y$ maps produced by the Planck
Collaboration in~2015~\cite{Planck2016ymap} and the Yang et
al.~\cite{Yang2007} galaxy group catalog extracted from SDSS Data
Release~7.  The group catalog contains $316041$ galaxy groups (after
masking), which are identified using a modified friends-of-friends algorithm that
allows ``groups'' with only a single member to be identified.  The
redshifts of the groups are spectroscopically measured and the halo
masses are estimated through an iterative process in which the group
assignment and mass-to-light ratio are updated at each step assuming
that the galaxies follow a Navarro-Frenk-White~\citep{Navarro1996} distribution around
each group center.  The Planck Collaboration has released two
component-separated Compton-$y$ maps, one derived from the Needlet
Internal Linear Combination (NILC) method, and one derived from the
Modified Internal Linear Combination Algorithm (MILCA).  We measure
the $y$-group correlation using both maps, finding generally
consistent results.

Our procedure for measuring the $y$-group cross-correlation is the
same as that described in \citet{Vikram2017} and we refer the reader
to that work for more details (including tests for foreground contamination
in the $y$ maps).  Briefly, we use
\texttt{treecorr}~\citep{Jarvis2004} to perform the $y$-group
cross-correlation measurement as a function of transverse radius in
six distinct group mass bins, using the same binning scheme as
\citet{Vikram2017}.  The covariance matrix of the measurements,
$\hat{C}_{y,g}^{ij}$ (where $i,j$ label each radial bin), is
determined using a spatial jackknife with~100 regions.  The measured
cross-correlation functions for both the MILCA and NILC $y$ maps are
shown in Fig.~\ref{fig:data}.  The results are nearly identical to
those presented in~\citet{Vikram2017}.  We fit
theoretical models to these measurements in \S\ref{sec:analysis}.

We also revisit the datasets used in tSZ analyses presented by~\citetalias{PlanckLBG2013} and Greco et al.~\cite{Greco2015}.  These
studies used a stacking method to constrain the Compton-$y$ signal
from halos over a wide mass range.  A set of ``isolation criteria''
were applied to galaxies selected from SDSS in order to minimize tSZ
signal arising from any gas not associated with the halo of interest.
The subsequent modeling and interpretation assumed that no such
unassociated gas was present in the data.  To illustrate how such a
measurement is susceptible to contamination from the two-halo term, in
\S\ref{sec:LBGs} we also measure the correlation between the Planck
$y$ maps and catalogs of locally brightest galaxies (LBGs) similar to
the catalog selected by \citetalias{PlanckLBG2013}.  The catalogs used for
this purpose were generated from the New York University Value-added
Catalog \citep{Blanton2005}. LBGs are defined to be those galaxies
with $z > 0.03$ that are brighter in $r$-band magnitude than all other
galaxies within a projected distance $R_{\rm iso}$ and within $|c
\Delta z_{\rm iso}|$.  The fiducial values for these criteria are
$R_{\rm iso} = 1$ Mpc/$h$ and $|c \Delta z_{\rm iso}| = 1000\,{\rm
  km}/{\rm s}$, but we will explore several choices.  Additional photometric
SDSS data are used to further remove any galaxies that could violate
the isolation criteria.  The construction
of these catalogs is described in more detail in \citet{Greco2015}.
We measure the LBG-$y$ cross-correlation using the same approach as
employed to measure the $y$-group correlation described above.  We
note that this is somewhat different than the analyses performed in
\citetalias{PlanckLBG2013} and~\citet{Greco2015}, which relied on matched-filter or
aperture photometry stacking analyses, using the multifrequency Planck
data to separate the tSZ signal from dust emission and other
contaminants.  In our analysis, the component separation has already
been performed on the full sky to produce the Planck $y$ maps.

\section{Halo Model}
\label{sec:halo}
\subsection{Fiducial Model}
\label{subsec:fid}
In order to interpret the $y$-group cross-correlation measurements, we make use of the halo models described in \citet{Vikram2017} (see also, e.g., \cite{Komatsu2002,Li2010,Fang2011}).  The fiducial pressure profile model underlying these calculations is that of~\citet{Battaglia2012}, who provide a fitting function to the results of their hydrodynamical simulations~\cite{Battaglia2010}.  This model predicts a relation $Y \propto M^{1.72}$, i.e., a power-law somewhat steeper than the self-similar value (see~\cite{Battaglia2012YM} for a full discussion of the $Y$--$M$ relation in these simulations).  Note that because the Battaglia pressure profile model is determined directly from cosmological hydrodynamics simulations, no explicit specification of the ``hydrostatic mass bias'' (often written as $(1-b)$ in the literature) is required.  As a rough estimate to provide context, we note that comparing the Battaglia pressure profile to the~\citet{Arnaud2010} pressure profile for the  massive, low-redshift population of clusters studied in the latter analysis yields a hydrostatic mass bias of roughly 10--15\% (i.e., $(1-b) \approx 0.85$--$0.9$).  However, the exact value varies with cluster-centric radius --- see Fig.~2 of~\citet{Battaglia2010}.  Below, we consider variations around the Battaglia model to test for evidence of departures from the fiducial pressure profile behavior.  A particular goal is to consider models in which the hot gas content of low-mass halos ($M \lesssim 10^{13.5} M_{\odot}$) --- which are unresolved in the \citet{Battaglia2012} simulations --- is suppressed or modified as a result of AGN and/or supernova feedback.  Note that our aim is thus not to calibrate the overall amplitude of the $Y$--$M$ relation, which is achieved most precisely via weak lensing observations of tSZ-selected cluster samples, but rather to constrain the mass dependence of this relation.  Specifically, we aim to test whether the mass dependence shows evidence for departures from the self-similar prediction ($Y \propto M^{1.67}$).  In Appendix~\ref{app:P0}, we consider an extended analysis in which the overall normalization of the pressure--mass relation ($P_0$) is allowed to vary, which yields constraints on the mass dependence generally consistent with those presented in our fiducial analysis.

We refer the reader to \citet{Vikram2017} for details of the halo modeling, and give only a brief summary of the theoretical framework here.  Note that our approach includes both the one-halo and two-halo contributions to the $y$-group correlation, whereas earlier stacking analyses have neglected the two-halo contribution (e.g.,~\cite{PlanckLBG2013,Greco2015,Crichton2016,Ruan2015,Spacek2016,Verdier2016}).\footnote{The modeling approach here is directly analogous to the modeling of the excess surface mass density $\Delta \Sigma$ in stacked weak lensing measurements.}  As shown in~\citet{Vikram2017}, however, the two-halo term can dominate the measured signal around low-mass halos, and thus cannot be neglected.

In the halo model, the excess Compton-\y parameter around a halo of mass $M$ at redshift $z$ is given by:
\beqa
\xi_{h,y}(r_\perp|M,z) = \frac{\sigma_T}{m_e c^2} \int_{-\infty}^{\infty} \frac{d\chi}{1+z} \xi_{h, P}(\sqrt{\chi^2 + r_\perp^2}|M,z).
\label{eq:yprof}
\eeqa
Here, $\sigma_T$ is the Thomson scattering cross section, $m_e c^2$ is the rest mass energy of an electron, $\chi$ and $r_\perp$ are line-of-sight and transverse co-moving distances, respectively, and $\xi_{h,P}(r|M,z)$ is the halo-pressure correlation function for halos of mass $M$ at redshift $z$.\footnote{Unless explicitly stated otherwise, the halo mass $M$ throughout is taken to be $M_{200}$, the mass contained within a halo-centric radius $r_{200}$ within which the mean enclosed density is $200$ times the critical density at the halo redshift.} The halo-pressure correlation function has a one-halo term from hot gas in the halo on which one is stacking, as well as a two-halo contribution from correlated neighboring systems:
\beqa
\xi^{\rm tot}_{h,P}(r|M,z) = \xi^{\rm one-halo}_{h,P}(r|M,z) + \xi^{\rm two-halo}_{h,P}(r|M,z).
\label{eq:xi_tot}
\eeqa
The one-halo term is simply:
\beqa
\xi^{\rm one-halo}_{h,P}(r|M,z) = P_e(r|M,z),
\label{eq:xi_oneh}
\eeqa
where $P_e(r|M,z)$ is the electron pressure at a co-moving distance $r$ from the halo of interest. 
The fiducial model in \citet{Vikram2017}, which we also adopt here, utilizes the fitting formulas from \citet{Battaglia2012} for the electron pressure profiles $P_e(r|M,z)$. In the analysis below, we will consider additional models, as we discuss subsequently.

The two-halo contribution to the halo-pressure correlation function is the Fourier-transform of the halo-pressure cross-power spectrum:
\beqa
\xi^{\rm two-halo}_{h,P}(r|M,z) = \int_0^{\infty} \frac{dk}{2\pi^2} k^2 \frac{{\rm sin}(kr)}{kr} P_{h,P}(k|M,z),
\label{eq:xi_twoh}
\eeqa
where the halo-pressure power spectrum is computed assuming linear halo bias as:
\beqa
P_{h,P}(k|M,z) = b(M) P_{\rm lin}(k) \int_0^{\infty} dM' \frac{dn}{dM'} b(M') u_P(k|M').
\label{eq:two-halo_power}
\eeqa
(For brevity of notation, we have suppressed the redshift labels in the right hand side of the equation.) Here, $M$ refers to the mass
of the halo on which one is stacking, while the integral over $M'$ describes the impact of correlated neighboring halos. 
In addition, $P_{\rm lin}(k)$ is the linear theory matter power spectrum, $\frac{dn}{dM}$ is the halo mass function, and $b(M)$ is the linear halo bias factor. 
The quantity $u_P(k|M')$ is
the Fourier transform of the pressure profile around a halo of mass $M'$:
\beqa
u_P(k|M') = \int_0^{\infty} dr \, 4 \pi r^2 \frac{{\rm sin}(kr)}{kr} P_e(r|M').
\label{eq:press_fourier}
\eeqa
Given a model for the electron pressure profile, $P_e(r|M')$, we can then compute the excess Compton-\y parameter around
halos of mass $M$ at redshift $z$ using Eqs.~\ref{eq:yprof}--\ref{eq:press_fourier}. 

Ultimately, the observationally accessible quantity of interest is the average excess Compton-\y parameter around groups in various mass bins, smoothed at the angular resolution
of Planck's component-separated \y map. Several additional steps are required to calculate this quantity from the halo-\y correlation function calculations outlined above.
First, we model the relation between the halo mass estimates of the \citet{Yang2007}
groups and their true underlying halo masses, allowing also for miscentering errors in their estimated positions, which act to suppress the observed signal. Second, we integrate over the redshift distribution of the groups, assuming Limber's
approximation to compute the resulting correlation function. Third, we smooth the model $y$-group cross-correlation function with a FWHM=10 arcminute beam to account for the resolution of the Planck
Compton-\y maps.  We label the smoothed $y$-group correlation function at transverse co-moving separation $r$ (we use this notation in what follows rather than $r_\perp$ for brevity) as
$\xi^s_{y,g}(r)$. 

We refer the reader to \citet{Vikram2017} for a detailed description of these steps. For the present purposes, we may summarize these procedures by noting that the fiducial model from \citet{Vikram2017} assumes that a lognormal distribution relates the estimated halo mass (referred to hereafter as the ``group mass'') and the true halo mass with a scatter of: $0.25$ dex in the two lowest group mass bins, $0.40$ dex in the middle three mass bins, and $0.30$ dex in the highest mass bin (see Fig.~\ref{fig:data} for the mass bin definitions). This fiducial model further assumes that the group mass lies, on average, $10\%$ above the true halo mass, as would arise if the~\citet{Yang2007} group finder has a small, yet systematic, tendency to accidentally include interloping foreground or background galaxies among the identified groups.  In order to model miscentering errors, the fiducial model assumes that a fraction $p_c$ of the \citet{Yang2007} groups are positioned precisely at the center of their host halos, while $1-p_c$ are miscentered according to the offset distribution and fractions of \citet{Johnston2007}.  As in~\citet{Vikram2017}, the fiducial values of the correctly-centered fraction in each mass bin are (from lowest to highest mass bin): $[0.53, 0.54, 0.58, 0.63, 0.72, 0.83]$.\footnote{Very recently, a weak lensing analysis of the~\citet{Yang2007} groups has appeared~\cite{Luo2017}; our fiducial assumptions about the sample are generally consistent with their results.}  In what follows, we vary both the correctly-centered fraction, $p_c^I$, and the average mass bias, $b_M^I$, in each mass bin (labeled by index $I$). This is important for testing the impact of uncertainties in the group catalog on our conclusions about the electron pressure profile.

\subsection{Alternative Models}
\label{subsec:alt}

We explore three variations around the fiducial \citet{Battaglia2012} pressure profile:
\begin{itemize}
\item {\bf Power law in mass ({\it PL})} In this case, we multiply the Battaglia pressure profile $P_e(r|M,z)$ by
an overall power law in halo mass $\propto (M/M_0)^{\alpha_{pl}}$, while fixing the pressure at
$M_0 \equiv 10^{14} M_\odot$:
\beqa
P_e(r|M,z) \rightarrow P_e(r|M,z) \left( \frac{M}{M_0} \right)^{\alpha_{pl}} \,.
\label{eq:PL}
\eeqa
Note that this variation impacts both the one-halo and two-halo terms (Eqs.~\ref{eq:xi_oneh} and~\ref{eq:xi_twoh}).  The self-similar case corresponds to $\alpha_{pl} = -0.05$, since the Battaglia model predicts $Y \propto M^{1.72}$.

\item {\bf Uncompensated break ({\it UB})} We also consider models in which hot gas is depleted
in halos below $M_0 = 10^{14} M_\odot$. This is meant to reflect the plausible impact of
AGN and/or supernova feedback (as motivated by, e.g.,~\cite{LeBrun2015,LeBrun2017,vandeVoort2016}) . In this case, we assume the Battaglia pressure profile at $M \geq M_0$ and multiply the pressure profile by $(M/M_0)^{\alpha_{ub}}$ for lower mass halos:\footnote{Note that the power-law Battaglia prediction is close enough to the self-similar model that the break models introduced here are effectively testing departures from either at low masses.}
\beqa
P_e(r|M,z) \rightarrow
\begin{cases}
P_e(r|M,z) \,, & M \geq M_0 \\
P_e(r|M,z) \left( \frac{M}{M_0} \right)^{\alpha_{ub}} \,, & M < M_0 \,.
\end{cases}
\label{eq:UB}
\eeqa
Here $\alpha_{ub} \geq 0$ hence corresponds to a thermal pressure suppression in halos below the break mass $M_0$.  This affects both the one-halo and two-halo terms  (Eqs.~\ref{eq:xi_oneh} and~\ref{eq:xi_twoh}) in general, though the contributions above the break mass are unchanged from the fiducial Battaglia model.  Note that the mean Compton-$y$ of the universe, $\langle y \rangle$, is not conserved in this model, as compared to the Battaglia pressure profile prediction (the same is true for the \textit{PL} model).  This observable presents another avenue for constraining gas pressure profiles and feedback processes~(e.g.,~\cite{Hill2015,Abitbol2017}).

\item {\bf Compensated break ({\it CB})} In a variant of the above model, we follow~\citet{Horowitz2017}
and assume that the ``suppressed'' portion of the gas in low-mass halos (following $(M/M_0)^{\alpha_{cb}}$) is
pushed out to large halo-centric radius, rather than removed from the host halo entirely. The suppressed gas is assumed to follow
the Gaussian distribution described in \cite{Horowitz2017}:\footnote{In \citet{Horowitz2017}, the width of this Gaussian distribution is taken to be $4 r_{\rm vir}$, while
here we adopt $2 r_{\rm vir}$ for the width. Our main conclusions should be insensitive to this choice.}
\beqa
P_e(r|M,z) \rightarrow
\begin{cases}
P_e(r|M,z) \,, & M \geq M_0 \\
P_e(r|M,z) \left( \frac{M}{M_0} \right)^{\alpha_{cb}} + A(\alpha_{cb}|M,z) \, e^{\frac{-r^2}{2(2r_{\rm vir})^2}} \,, & M < M_0 \,,
\end{cases}
\label{eq:CB}
\eeqa
where $A(\alpha_{cb}|M,z)$ is a normalization parameter determined such that the total thermal energy (i.e., integrated pressure) of each halo is fixed. This model varies the one-halo term (Eq.~\ref{eq:xi_oneh}) around the Battaglia form at sub-break masses, while fixing the two-halo contribution at large separations. This is because the latter quantity depends only on the total thermal energy of each halo and not the precise pressure distribution (Eq.~\ref{eq:xi_twoh}).  Note that by construction $\langle y \rangle$ is conserved in this model, as compared to the Battaglia pressure profile prediction.
\end{itemize}

\section{Analysis}
\label{sec:analysis}
\subsection{Thermal SZ -- Galaxy Group Cross-Correlation Function}
\label{sec:fitting}
\subsubsection{Pressure Profile Constraints}

Here we present the results of fitting the theoretical models described in the previous section to the $y$-group cross-correlation function measured using the MILCA/NILC $y$ maps~\cite{Planck2016ymap} and the \citet{Yang2007} SDSS DR7 group catalog.  For simplicity, we discard the lowest mass bin in our fitting analysis, as we find that it contributes negligibly to the final constraints, while possessing the largest off-diagonal covariances with the other mass bins (as estimated via jackknife in \S\ref{sec:data}).  We include the off-diagonal covariance matrix elements between radial bins within each mass bin, but discard the off-diagonal blocks between different mass bins.  In terms of the correlation matrix $\hat{\rho}_{y,g}^{ij} \equiv \hat{C}_{y,g}^{ij}/\sqrt{\hat{C}^{ii}_{y,g} \hat{C}^{jj}_{y,g}}$, we find that typical values in the off-diagonal blocks between different mass bins are $\lesssim 25$\%, with significant noise fluctuations due to the relatively small number of jackknife regions.  Including these blocks in the likelihood calculation yields best-fit parameters consistent with those found when neglecting them below, but the parameter error estimation is significantly more robust when the block-diagonal covariance matrix is used.  We thus adopt this approach in the following.

\begin{figure}
\centering
\includegraphics[width=\textwidth]{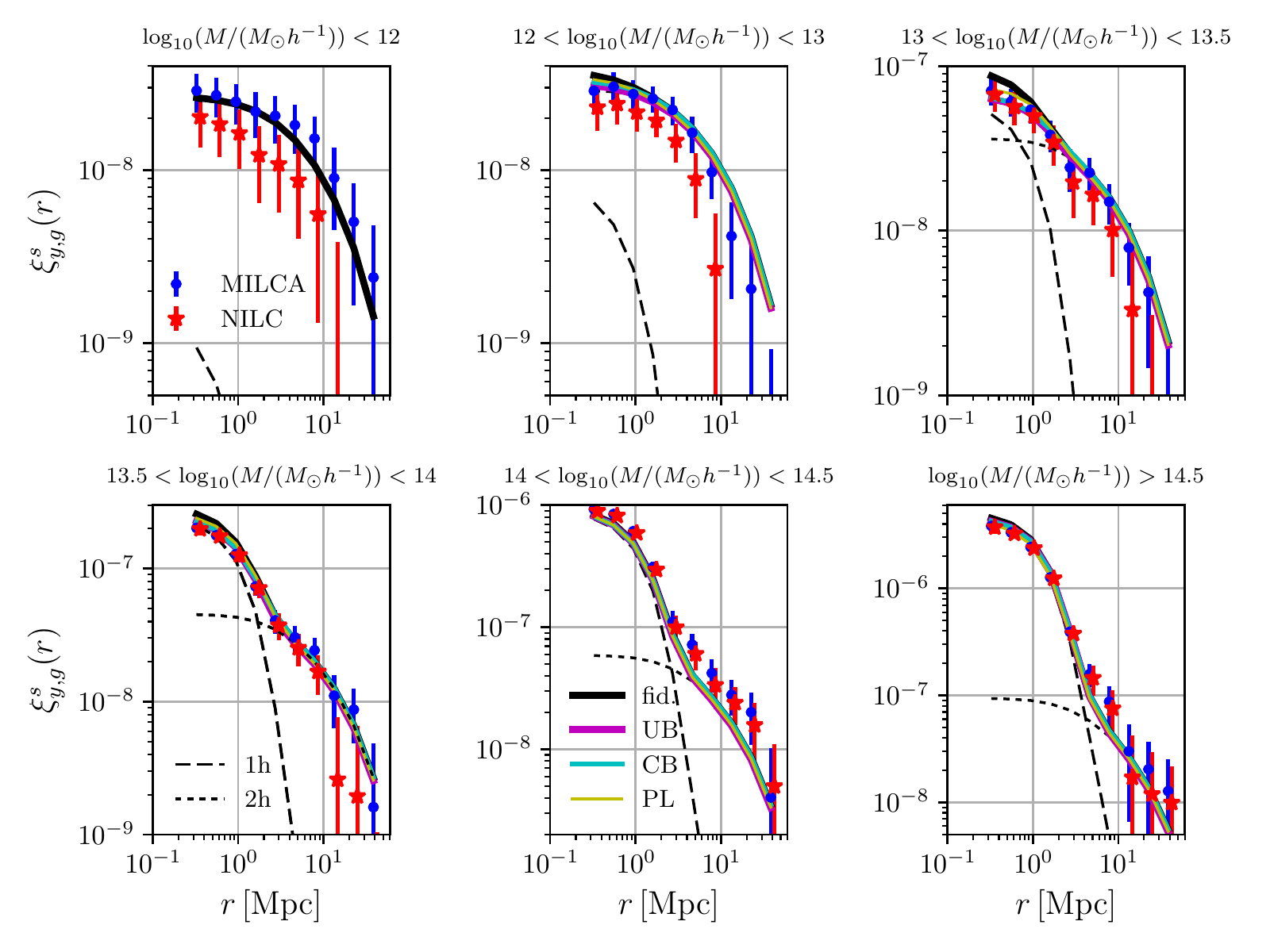}
\caption{\label{fig:data} Compton-$y$--group cross-correlation function measurements and theoretical models.  The blue circles (red stars) show the measurements for the MILCA (NILC) $y$ map, with error bars computed from the diagonal elements of the covariance matrix.  The black curves are the fiducial Battaglia pressure profile model, with the one-halo and two-halo contributions shown in long-dashed and short-dashed, respectively.  The two-halo term completely dominates the signal in the two lowest mass bins, so it is hard to distinguish the short-dashed and solid curves in these bins.  The other curves correspond to the best-fit results (MILCA data only) for the three model variants described in \S\ref{sec:halo}: the {\it UB} model (magenta), the {\it CB} model (cyan), and {\it PL} model (yellow).  The best-fit results for each model lie very close to one another, making them hard to distinguish by eye on the plot.  Note that the best-fit NILC results are similar to those for MILCA, as seen in Table~\ref{tab:alpha}.  The lowest mass bin is not used in the fitting analysis, and thus we do not plot best-fit models for that bin.  The underlying $Y$--$M$ relations inferred from these fits are shown in Fig.~\ref{fig:YM}.}
\end{figure}

For each of the pressure profile models considered, there is a free parameter ($\alpha_{pl}$, $\alpha_{ub}$, or $\alpha_{cb}$) associated with the mass-dependence of the gas pressure profile.  Priors on these parameters are described below.  In addition, each of the five mass bins considered (labeled by $I,J$) has a free parameter $p_c^I$ characterizing the fraction of correctly centered halos in this bin.  For each model, we define a Gaussian likelihood function (parameter arguments suppressed for brevity):
\beqa
-2 \, {\rm ln} \, \mathcal{L} \equiv \chi^2 = \sum_{i,j} \left( \hat{\xi}^{s,i}_{y,g} - \xi^{s,i}_{y,g} \right) \left[ \hat{C}_{y,g}^{-1} \right]^{ij} \left( \hat{\xi}^{s,j}_{y,g} - \xi^{s,j}_{y,g} \right) \,,
\label{eq:likelihood}
\eeqa
where $\hat{\xi}^{s,i}_{y,g}$ is the measured cross-correlation function in the $i^{\rm th}$ bin, $\xi^{s,i}_{y,g}$ is the theoretical prediction in this bin (appropriately averaged over the radial extent of the bin), and $\hat{C}^{ij}_{y,g}$ is the block-diagonal covariance matrix described above.\footnote{Note that we neglect the parameter dependence of the covariance matrix, as it is dominated by contributions from noise due to the CMB, instrument, and residual foregrounds.}  Note that $\alpha$ is the only parameter in the likelihood that affects multiple mass bins; the correctly-centered fraction $p_c^I$ affects only the $I^{\rm th}$ mass bin.  In combination with the block-diagonal nature of the covariance, this allows the full likelihood to be computed from the set of two-parameter likelihoods for $(\alpha, p_c^I)$ constructed for each mass bin.  Thus, we can rapidly compute the full likelihood function, and do not require Monte Carlo Markov Chain techniques to obtain parameter constraints.

\begin{table}[ht]
\begin{center}
  MILCA \vspace{2pt} \\
  \begin{tabular}{| c | c | c | c | c |}
    \hline 
     Model & Parameter [prior range] & Marginalized constraint & Global best-fit [$p_c^I$] & $\chi^2$ [$\Delta \chi^2_{\rm fid}$] \\
     \hline \hline
    {\it PL} & $\alpha_{pl}$ [-1, 1] & $-0.05\pm0.04$ & -0.05 [0.22,0.001,0.44,0.72,0.69] & 61.0 [20.2] \\ \hline
    {\it UB} & $\alpha_{ub}$ [-1, 1.25] & $0.34^{+0.20}_{-0.19}$ & 0.17 [0.43,0.06,0.55,0.74,0.66] & 61.3 [19.9] \\ \hline
    {\it CB} & $\alpha_{cb}$ [0, 2] & $0.66\pm0.34$ & 0.36 [0.61,0.16,0.65,0.77,0.66] & 60.5 [20.7] \\ \hline
  \end{tabular}
  \\ \vspace{12pt} NILC \vspace{2pt} \\
    \begin{tabular}{| c | c | c | c | c |}
    \hline 
     Model & Parameter [prior range] & Marginalized constraint & Global best-fit [$p_c^I$] & $\chi^2$ [$\Delta \chi^2_{\rm fid}$] \\
     \hline \hline
    {\it PL} & $\alpha_{pl}$ [-1, 1] & $-0.08\pm0.04$ & -0.08 [0.21,0.11,0.47,0.66,0.68] & 68.4 [32.4] \\ \hline
    {\it UB} & $\alpha_{ub}$ [-1, 1.25] & $0.49^{+0.23}_{-0.22}$ & 0.37 [0.10,0.41,0.70,0.69,0.61] & 68.3 [32.5] \\ \hline
    {\it CB} & $\alpha_{cb}$ [0, 2] & $0.57\pm0.29$ & 0.38 [0.81,0.43,0.71,0.69,0.60] & 71.4 [29.4] \\ \hline
  \end{tabular}
  \caption{Constraints on the mass dependence of the electron pressure profile for various theoretical models (see \S\ref{sec:halo} for model and parameter definitions).  The fiducial Battaglia model in all cases corresponds to $\alpha_{pl} = \alpha_{ub} = \alpha_{cb} = 0$, with $\chi^2 = 81.2$ (MILCA) and $\chi^2 = 100.8$ (NILC).  The third column gives constraints after marginalizing over the correctly-centered fraction of halos in each of the five mass bins, with an uninformative prior on the centered fraction $p_c^I \in [0,1]$ for all bins.  The quoted values are the mean and $68\%$ C.L. intervals computed from the marginalized posterior.  The fourth column gives the global best-fit point, including the centered fraction values in brackets (from the second-lowest mass bin to highest mass bin), although these are not individually well-constrained.  The fifth column gives the $\chi^2$ values associated with the global best-fit for each case, as well as the improvement in $\chi^2$ with respect to the fiducial model.  The fiducial and best-fit models (MILCA-only) are plotted in Fig.~\ref{fig:data}.}
  \label{tab:alpha}
\end{center}
\end{table}

In general, the correctly-centered fractions $p_c^I$ are not well-constrained by the data, so we adopt uninformative priors ($p_c^I \in [0,1]$) on these parameters and marginalize over them in order to obtain constraints on $\alpha_{pl}$, $\alpha_{ub}$, and $\alpha_{cb}$.  We also use the Nelder-Mead algorithm to find global best-fit points in the full parameter space for each model, which are plotted in Fig.~\ref{fig:data}.

The results of the analysis for each pressure profile model are given in Table~\ref{tab:alpha} and summarized in the following:
\begin{itemize}
\item {\bf Power law in mass ({\it PL})} For this model, we adopt a flat prior $\alpha_{pl} \in [-1,1]$, centered on the fiducial value $\alpha_{pl} = 0$.  Marginalizing over the correctly-centered fractions $p_c^I$, the posterior yields constraints that are consistent with $\alpha_{pl} = 0$ at $\approx 1\sigma$ for MILCA, with a $\approx 2\sigma$ indication of $\alpha_{pl} < 0$ for NILC.  These results are consistent with self-similarity ($\alpha_{pl} = -0.05$) and with the result obtained for the same model in \citet{Greco2015}, who used a different component-separation and stacking approach.

\item {\bf Uncompensated break ({\it UB})} For this model, we assume a flat prior $\alpha_{ub} \in [-1,1.25]$.  The marginalized constraints on $\alpha_{ub}$ indicate a $\approx 2\sigma$ preference for $\alpha_{ub} > 0$, corresponding to a suppression of the electron pressure in halos below the break mass $M_0 \equiv 10^{14} M_{\odot}$, with NILC preferring a slightly larger suppression than MILCA.

\item {\bf Compensated break ({\it CB})} For this model, we implement a flat prior $\alpha_{cb} \in [0,2]$.  Note that the fiducial model ($\alpha_{cb} = 0$) lies at the edge of the prior range in this case, because values $\alpha_{cb} < 0$ yield unphysical negative pressure model predictions.  We test the effect of extending the prior range to $\alpha_{cb} < 0$ by assuming the \textit{UB} model for this range, and find that the marginalized constraints on $\alpha_{cb}$ are nearly unchanged.  Thus, we conclude that the prior is not strongly driving the results.  As for the \textit{UB} model, the marginalized constraints on $\alpha_{cb}$ yield a $\approx 2\sigma$ preference for $\alpha_{cb} > 0$, corresponding to a suppression of the electron pressure in halos below the break mass scale.  The MILCA and NILC results are very similar.
\end{itemize}

In all cases, the best-fit models are a significantly better fit than the fiducial Battaglia pressure profile model, as seen in the final column of Table~\ref{tab:alpha}.  However, there is not a strong preference for any of the particular pressure profile models (\textit{PL}, \textit{UB}, or \textit{CB}) --- the $\chi^2$ values for the best-fit parameter values in each case are very similar.  Higher-resolution, lower-noise data will be needed in order to determine whether the ``break'' models are preferred over the simple power-law model, as well as whether the data prefer $CB$-type models in which the total thermal energy is conserved or $UB$-type models in which it is not.

\begin{figure}[ht]
\centering
\begin{tabular}{cc}
\includegraphics[width=0.5\textwidth]{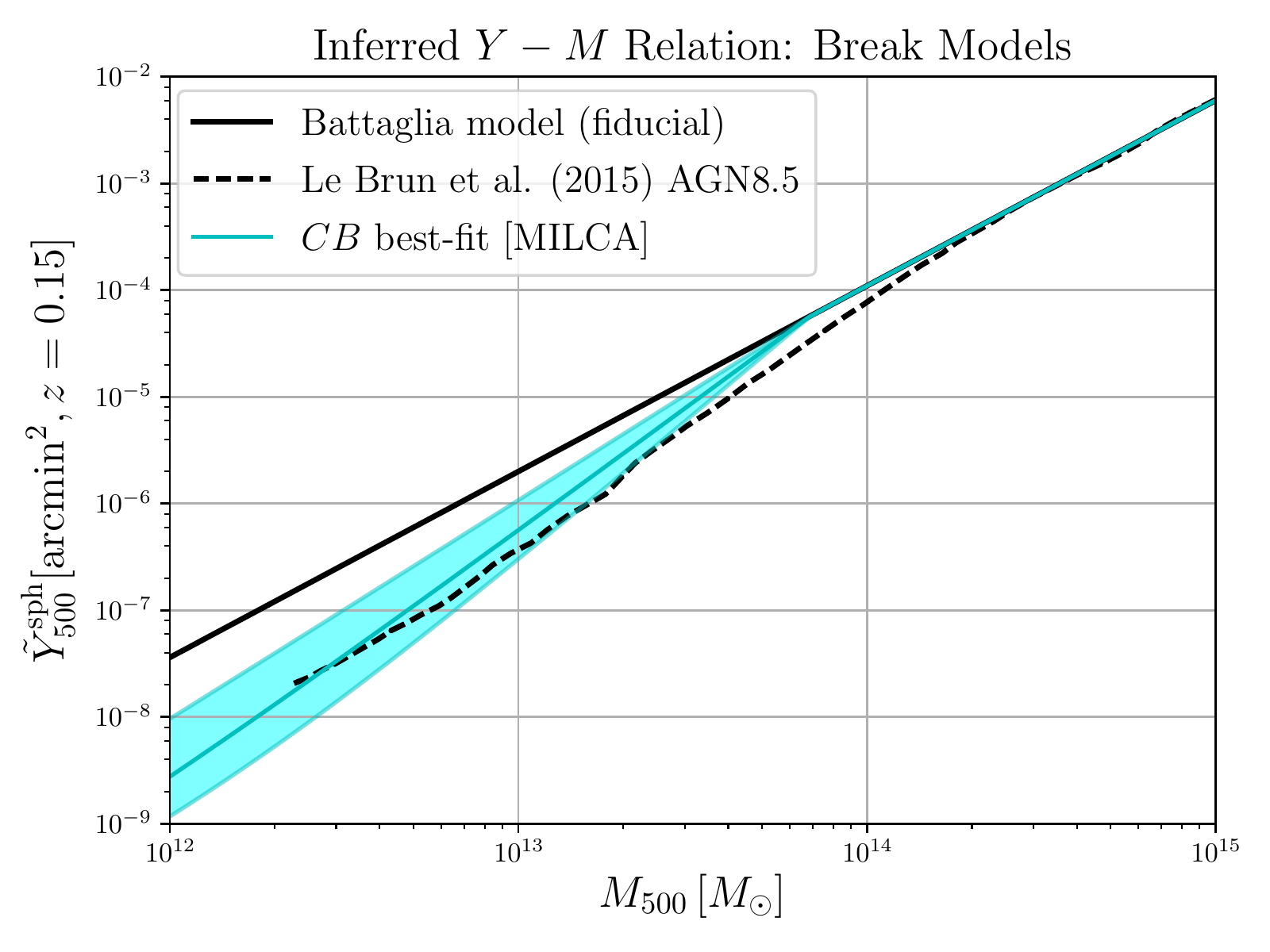} & \includegraphics[width=0.5\textwidth]{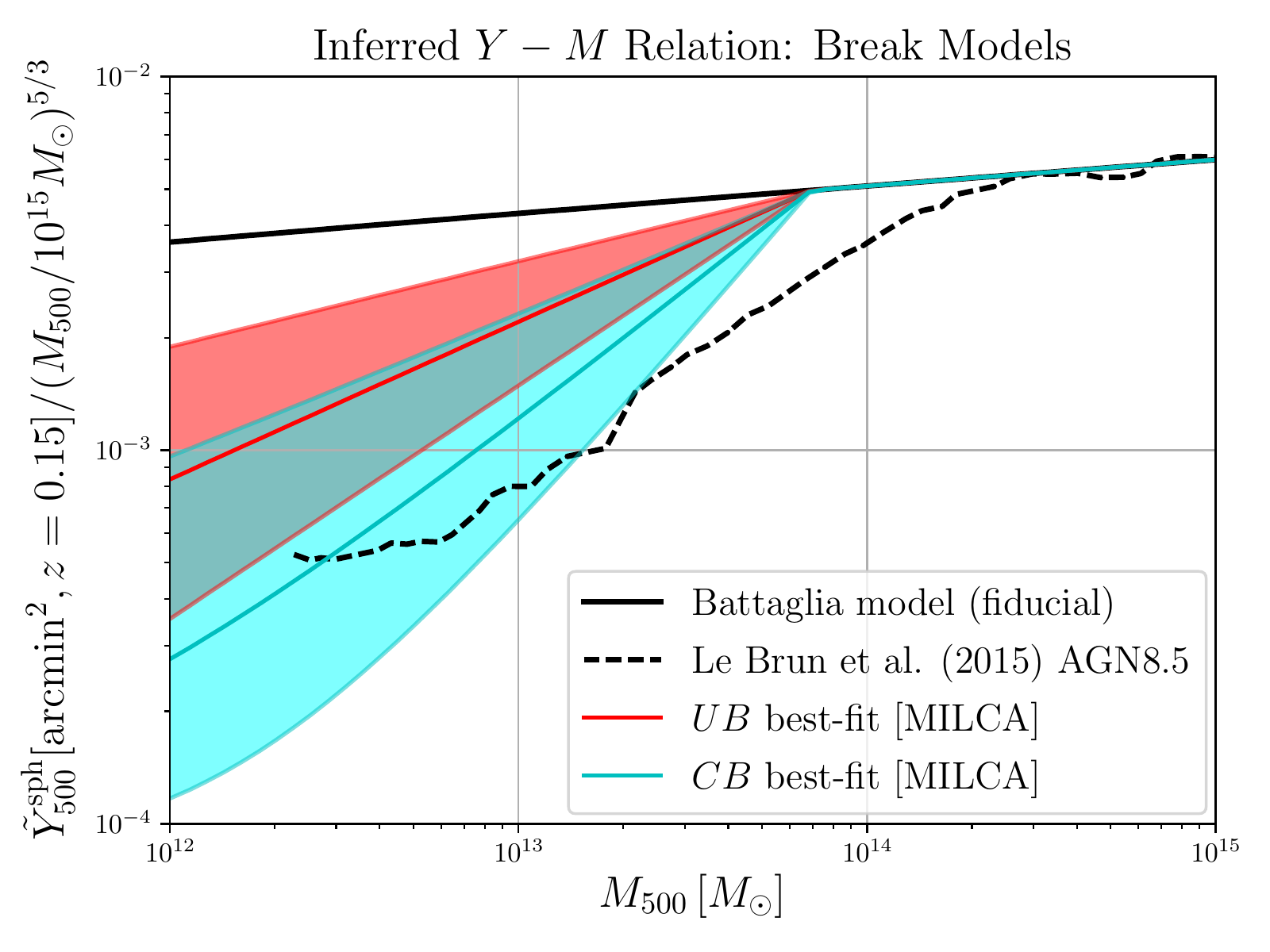} \\
\includegraphics[width=0.5\textwidth]{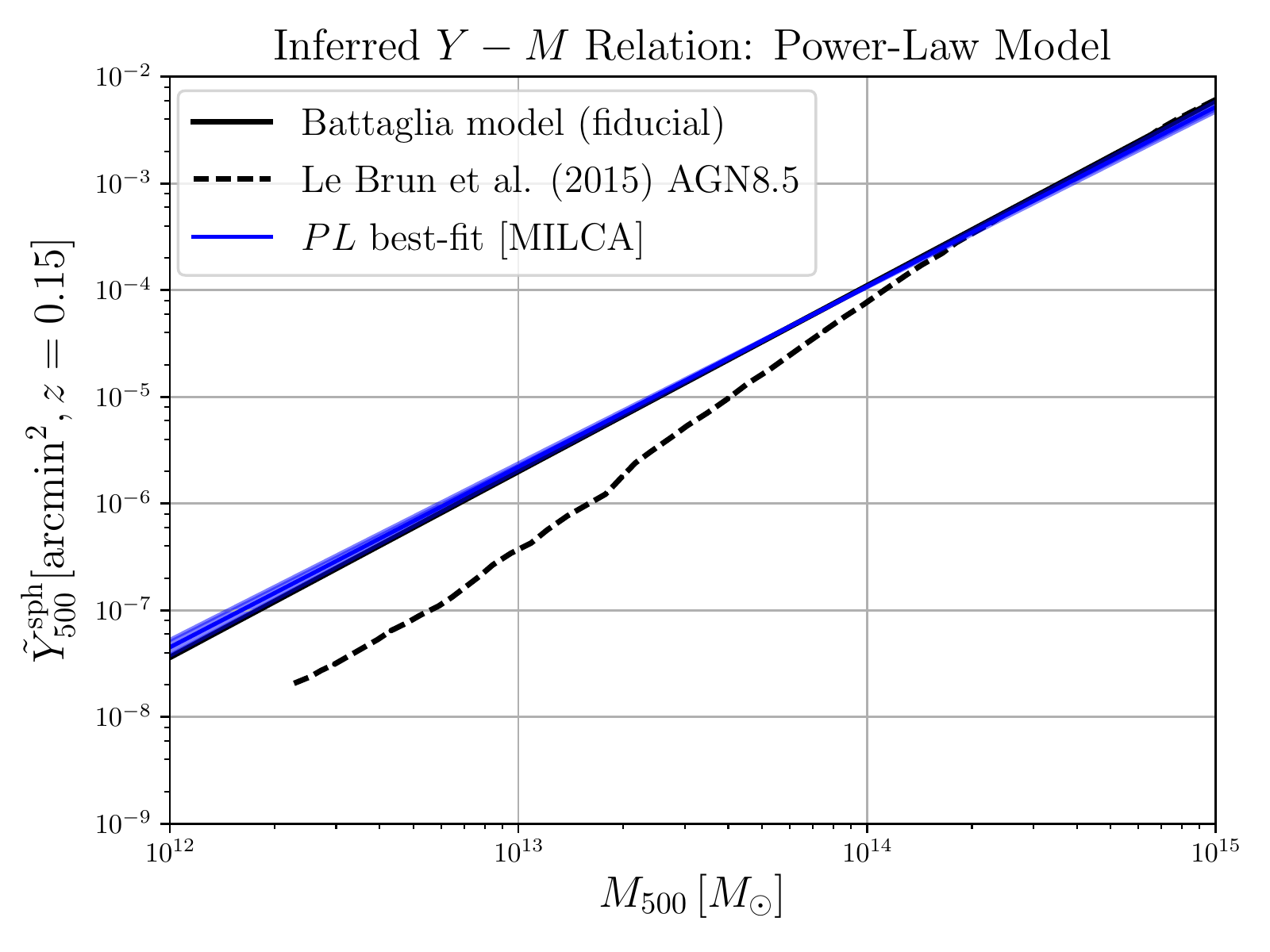} & \includegraphics[width=0.5\textwidth]{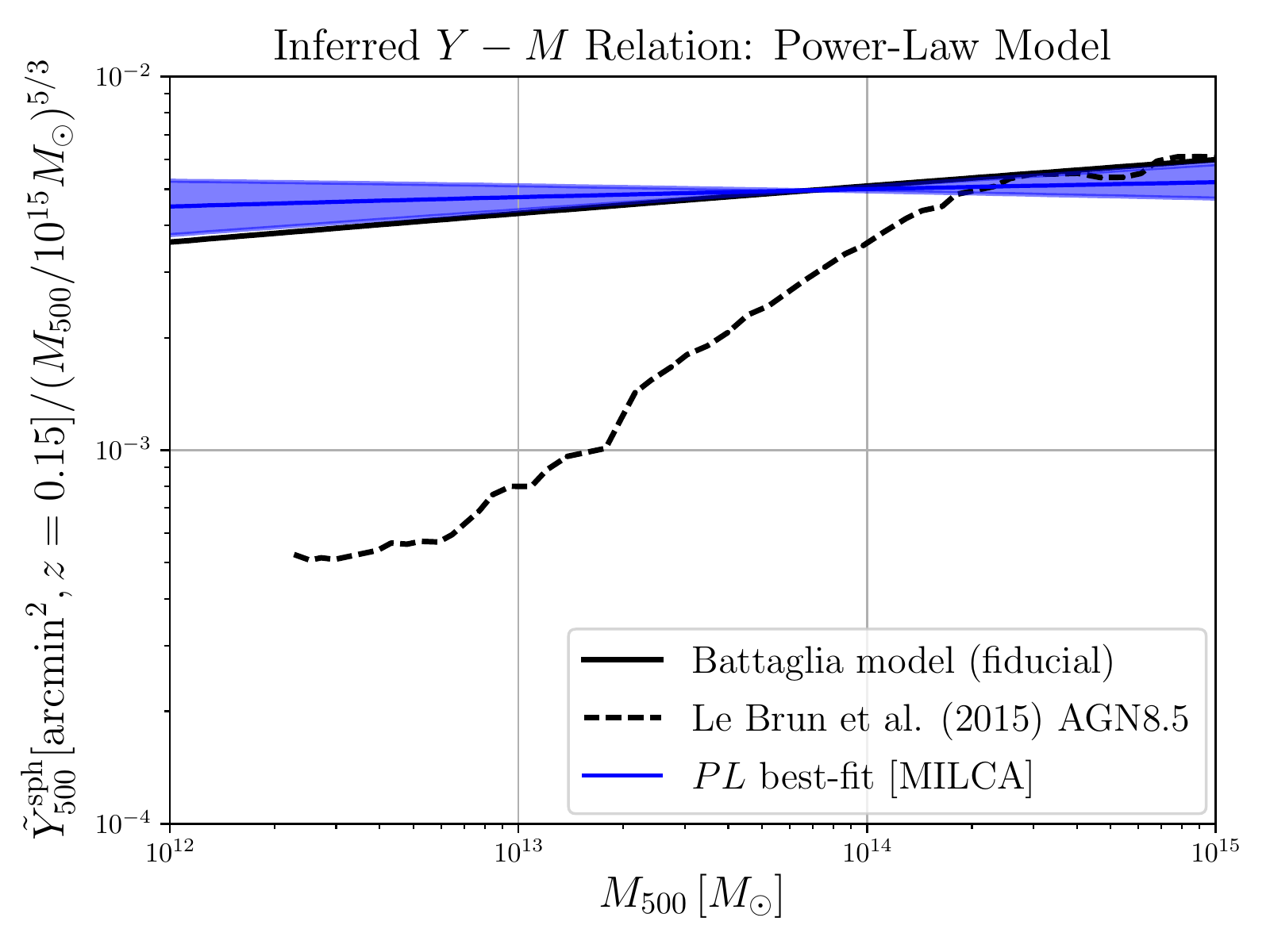}
\end{tabular}
\caption{\label{fig:YM} Inferred $\tilde{Y}^{\rm sph}_{500}$--$M_{500}$ relations based on fitting the theoretical models described in \S\ref{sec:halo} to the $y$--group cross-correlation function measurements shown in Fig.~\ref{fig:data}.  The top row shows results for the break models (\textit{UB} and \textit{CB}), while the bottom row shows results for the power-law (\textit{PL}) model.  The left panels show $\tilde{Y}^{\rm sph}_{500}$ as a function of $M_{500}$, while the right panels show $\tilde{Y}^{\rm sph}_{500} / (M_{500} / 10^{15} \, M_{\odot})^{5/3}$ as a function of $M_{500}$ (i.e., the scaling relation divided by the self-similar prediction).  For clarity, we show the MILCA $y$ map results only; the NILC $y$ map results are very similar, as seen in Table~\ref{tab:alpha}.  The solid black line shows the fiducial Battaglia pressure profile model, which has been extended to mass scales well below those where it was calibrated in simulations (roughly $M_{200} \gtrsim 5 \times 10^{13} M_{\odot}/h$)~\cite{Battaglia2012}.  The dashed black curve is the theoretical prediction of the ``AGN 8.5'' model from the hydrodynamical simulations of \citet{LeBrun2015}, as extracted directly from the simulated halo catalogs (i.e., no fitting function is used).  The cyan, red, and blue curves and shaded bands show the best-fit result and $1\sigma$ confidence region for the {\it CB}, {\it UB}, and {\it PL} models, respectively.  We present the break and power-law models in separate plots, as the \textit{PL} fit is driven by the high signal-to-noise measurements in the highest two mass bins ($M > 10^{14} \, M_{\odot}/h$), which play essentially no role in the \textit{UB}/\textit{CB} model fits, due to the models' parameterization.  We show only the \textit{CB} model in the upper left panel for clarity.  Note that the plots show $M_{500}$, while our calculations are in terms of $M_{200}$ (see \S\ref{sec:halo}), which is why the break in the {\it UB} and {\it CB} models occurs below $10^{14} \, M_{\odot}$ here.}
\end{figure}

For straightforward comparison with previous studies, we compute the $Y$--$M$ relation associated with each pressure profile model, with the results plotted in Fig.~\ref{fig:YM}.  In particular, we compute
\beqa
\tilde{Y}_{500}^{\rm sph}(M,z) \equiv \frac{(d_A(z) / 500 \, {\rm Mpc})^2}{E^{2/3}(z)} \frac{\sigma_T}{m_e c^2} \int_0^{r_{500}} 4\pi r^2 dr \, P_e(r | M,z) / d_A(z)^2 \,,
\label{eq:Y500}
\eeqa
where $E(z) \equiv H(z)/H_0$ is the dimensionless Hubble parameter and $r_{500}$ is the radius enclosing a mass within which the mean density is 500 times the critical density at redshift $z$.  We compute the results at a characteristic redshift $z = 0.15$, typical of the galaxy groups in the \citet{Yang2007} sample.  Fig.~\ref{fig:YM} shows the results for the fiducial Battaglia model, which has been extended to mass scales well below those where it was determined in the original simulations ($M_{200} \gtrsim 5 \times 10^{13} \, M_{\odot}/h$).  We also show the $p_c^I$-marginalized constraints for the \textit{PL}, \textit{UB}, and \textit{CB} models.  We plot the latter two models (i.e., the break models) separately from the power-law model, as the data driving the fits in each case are somewhat different.  In particular, the $\alpha_{ub}$ and $\alpha_{cb}$ fits receive essentially no information from the cross-correlation results for high-mass ($M > 10^{14} \, M_{\odot}/h$) objects, due to the models' parameterization.  These models explicitly probe the pressure behavior in low-mass systems: as seen in Fig.~\ref{fig:YM}, there is moderate evidence ($\approx 2 \sigma$) for a suppression in the electron pressure in low-mass groups.  In contrast, the $\alpha_{pl}$ fit is dominated by the high-signal-to-noise measurements for massive systems, which overcome the preference for a suppression in the low-mass data and lead to a preference for $\alpha_{pl}$ values near the self-similar value.  The \textit{PL} parameterization thus obscures the moderate evidence for pressure suppression seen in the \textit{UB} and \textit{CB} fits.  As expected based on previous studies~(e.g.,~\cite{Greco2015,Planck2013profile}), the $Y$--$M$ relation for massive objects is consistent with the Battaglia ($\alpha_{pl} = 0$) or self-similar ($\alpha_{pl} = -0.05$) predictions.

For comparison, Fig.~\ref{fig:YM} also shows the simulation predictions of the ``AGN 8.5'' model from~\citet{LeBrun2015} (scaled in redshift as in Eq.~\ref{eq:Y500}), which suggest a break near the mass scale assumed in our \textit{UB} and \textit{CB} models (note that their ``AGN 8.0'' model falls between the dashed and solid black curves in Fig.~\ref{fig:YM}, and predicts a break at a slightly lower mass than the AGN 8.5 model).  The AGN 8.5 data points are drawn directly from the simulated halo catalogs, rather than calculated from a fitting function (which is why small fluctuations are visible).  We note that although the feedback prescriptions in~\citet{Battaglia2012} and~\citet{LeBrun2015} yield similar predictions for most tSZ observables, the~\citet{LeBrun2015} simulation resolves smaller halos, where the power-law fitting function derived by~\citet{Battaglia2012} likely breaks down (also, note that the AGN 8.0 model lies closer to~\citet{Battaglia2012} than AGN 8.5).  The inferred relations for the \textit{UB} and \textit{CB} models are in general agreement with the AGN 8.5 prediction at low halo masses, although the uncertainties are relatively large.  Note that the {\it CB} model has a non-trivial shape at low masses for large values of $\alpha_{cb}$, as the compensated, Gaussian part of the pressure profile becomes significant within $r_{500}$ (see \S\ref{sec:halo}).

\subsubsection{Mass Bias Constraints}
In addition to the pressure profile model variations described above, we also consider a model in which the pressure profile is fixed to the fiducial Battaglia case, but the mass bias $b_M^I$ in each mass bin is allowed to vary.  In our fiducial model, the group mass is assumed to be 10\% higher than the true halo mass, i.e., $b_M^I = 0.1$ in each mass bin.  Here, we assume a flat prior $b_M^I \in [0,0.6]$ for the mass bias in each mass bin.  We also allow the correctly-centered fraction $p_c^I$ to vary in each bin, with an uninformative flat prior, as in the analyses above.  In this model, the likelihood for each mass bin is completely independent, as the parameters $(b_M^I, p_c^I)$ are independent for each bin.  For consistency with the pressure profile analyses above, we do not consider the lowest mass bin here.

The results of this analysis are given in Table~\ref{tab:bM}.  In general, the mass bias constraints are consistent at $\approx 1$--$2\sigma$ with the assumed value $b_M^I = 0.1$ in our fiducial model, although the third ($13.5 < \log_{10}(M / (M_{\odot} h^{-1})) < 14$) and fifth ($14.5 < \log_{10}(M / (M_{\odot} h^{-1}))$) bins prefer somewhat larger biases at $2$--$3 \sigma$.  The overall goodness-of-fit in this model is better than that found for any of the pressure profile model variations, with $\Delta \chi^2_{\rm fid} = 34.2$ for MILCA and $\Delta \chi^2_{\rm fid} = 47.7$ for NILC (compare to $\Delta \chi^2_{\rm fid}$ values in Table~\ref{tab:alpha}).  However, this comes at the cost of four additional parameters in comparison to the pressure profile models (ten vs. six).  In addition, the fifth bin (highest mass bin) is essentially insensitive to $\alpha_{ub}$ or $\alpha_{cb}$ (apart from small two-halo contributions), and thus does not contribute to the evidence for $\alpha_{ub} > 0$ and $\alpha_{cb} > 0$ found above. Overall, the data do not obviously prefer large mass biases instead of a suppression in the electron pressure profile.

\begin{table}[h]
\begin{center}
  MILCA: best-fit $\chi^2 = 47.0$ [$\Delta \chi^2_{\rm fid} = 34.2$] \vspace{2pt} \\
  \begin{tabular}{| c | c | c | c | c |}
    \hline 
     Mass Bin & Parameter [prior range] & Marginalized constraint & Best-fit [$p_c^I$] & $\chi^2$ \\
     \hline \hline
    $12 < \log_{10}(M / (M_{\odot} h^{-1})) < 13$ & $b_M^1$ [0, 0.6] & $0.33\pm0.19$ & 0.44 [0.48] & 12.1 \\ \hline
    $13 < \log_{10}(M / (M_{\odot} h^{-1})) < 13.5$ & $b_M^2$ [0, 0.6] & $0.32\pm0.16$ & 0.22 [0.04] & 5.7 \\ \hline
    $13.5 < \log_{10}(M / (M_{\odot} h^{-1})) < 14$ & $b_M^3$ [0, 0.6] & $0.30\pm0.08$ & 0.32 [0.78] & 5.4 \\ \hline
    $14 < \log_{10}(M / (M_{\odot} h^{-1})) < 14.5$ & $b_M^4$ [0, 0.6] & $0.05\pm0.04$ & 0.0 [0.61] & 17.6 \\ \hline
    $14.5 < \log_{10}(M / (M_{\odot} h^{-1}))$ & $b_M^5$ [0, 0.6] & $0.26\pm0.06$ & 0.25 [0.73] & 6.2 \\ \hline
  \end{tabular}
  \\ \vspace{12pt} NILC: best-fit $\chi^2 = 53.1$ [$\Delta \chi^2_{\rm fid} = 47.7$] \vspace{2pt} \\
    \begin{tabular}{| c | c | c | c | c |}
    \hline 
     Mass Bin & Parameter [prior range] & Marginalized constraint & Best-fit [$p_c^I$] & $\chi^2$ \\
     \hline \hline
    $12 < \log_{10}(M / (M_{\odot} h^{-1})) < 13$ & $b_M^1$ [0, 0.6] & $0.36^{+0.18}_{-0.19}$ & 0.60 [1.0] & 20.0 \\ \hline
    $13 < \log_{10}(M / (M_{\odot} h^{-1})) < 13.5$ & $b_M^2$ [0, 0.6] & $0.36^{+0.16}_{-0.15}$ & 0.34 [0.35] & 7.2 \\ \hline
    $13.5 < \log_{10}(M / (M_{\odot} h^{-1})) < 14$ & $b_M^3$ [0, 0.6] & $0.30^{+0.08}_{-0.07}$ & 0.32 [0.85] & 8.3 \\ \hline
    $14 < \log_{10}(M / (M_{\odot} h^{-1})) < 14.5$ & $b_M^4$ [0, 0.6] & $0.05\pm0.04$ & 0.0 [0.54] & 11.4 \\ \hline
    $14.5 < \log_{10}(M / (M_{\odot} h^{-1}))$ & $b_M^5$ [0, 0.6] & $0.28\pm0.06$ & 0.28 [0.71] & 6.2 \\ \hline
  \end{tabular}
  \caption{Constraints on the mass bias in each bin, assuming the
    fiducial Battaglia pressure profile model (see \S\ref{sec:halo}
    for model and parameter definitions).  The fiducial model assumes
    $b_M^I = 0.1$ in all bins, with $\chi^2 = 81.2$ (MILCA) and
    $\chi^2 = 100.8$ (NILC).  The third column gives constraints after
    marginalizing over the correctly-centered fraction of halos in
    each mass bin, with an uninformative prior on the centered
    fraction $p_c^I \in [0,1]$ for all bins.  The quoted values are
    the mean and $68\%$ C.L. intervals computed from the marginalized
    posterior.  The fourth column gives the best-fit point in the 2D
    parameter space for each mass bin, including the centered fraction
    values in brackets, although these are not as tightly constrained
    as the mass biases.  The fifth column gives the $\chi^2$ values
    associated with the best-fit point for each mass bin.  The overall
    $\chi^2$ and improvement with respect to the fiducial model
    ($\Delta \chi^2_{\rm fid}$) are given above each table.}
  \label{tab:bM}
\end{center}
\end{table}

Moreover, we can independently constrain possible mass biases by
measuring the group auto-correlation function in each mass bin,
$w(\theta)$, and comparing to theoretical models constructed
analogously to the approach described in \S\ref{sec:halo}.\footnote{See also
various tests presented in~\citet{Yang2007} that indicate that large
mass biases in the group catalog are unlikely.}  We focus
in particular on the second ($13 < \log_{10}(M / (M_{\odot} h^{-1})) <
13.5$) and third ($13.5 < \log_{10}(M / (M_{\odot} h^{-1})) < 14$)
mass bins, as these drive the constraints on $\alpha_{ub}$ and
$\alpha_{cb}$.  If there is a large mass bias in these bins, this
should also show up in $w(\theta)$, which depends quadratically on the halo clustering bias (which is itself mass-dependent); if
instead the $y$-group cross-correlation results are due to a
suppression of the electron pressure below $M_0$, the $w(\theta)$
results should be consistent with the fiducial mass bias.

\begin{figure}
\centering
\includegraphics[width=0.6\textwidth]{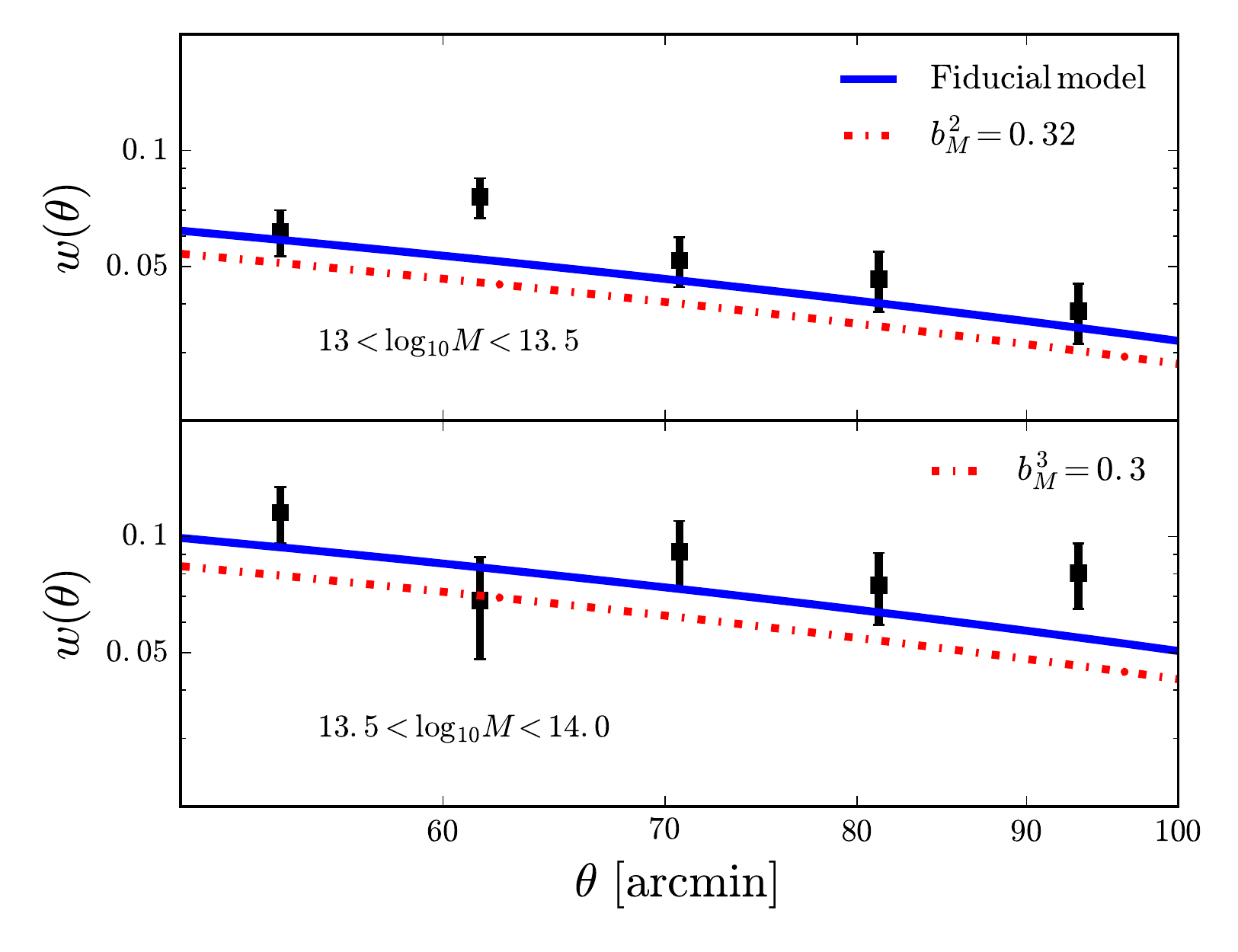}
\caption{\label{fig:wtheta} Measurement of $w(\theta)$ for groups with
  $13 < \log_{10} ( M / (M_{\odot}/h) ) < 13.5$ and $13.5 <
  \log_{10}(M / (M_{\odot} h^{-1})) < 14$, which are the mass bins
  that drive the constraints on $\alpha_{ub}$ and $\alpha_{cb}$ in
  Table~\ref{tab:alpha}.  The theory curves show our fiducial model
  (solid blue), which assumes a mass bias $b_M^I = 0.1$ in each mass
  bin, and a model that assumes the group masses overestimate the true
  halo masses by $32\%$ (top panel) or $30\%$ (bottom panel), 
  i.e., $b_M^I = 0.32$ or 0.30 (dashed red).  The latter curves are
  motivated by the results in Table~\ref{tab:bM}.  The
  high-bias model is disfavored relative the fiducial model by
  $1.9\sigma$.  Thus, it is unlikely that the evidence seen earlier
  for a suppression of the electron pressure in low-mass groups is
  entirely due to unaccounted-for mass biases.}
\end{figure}

The measurements of $w(\theta)$ for these two bins are shown in
Fig.~\ref{fig:wtheta}.  We also show two theoretical models:
$w(\theta)$ calculated assuming the fiducial bias $b_M^I = 0.1$, and
$w(\theta)$ calculated assuming the mean posterior values of $b_M^I$
from Table~\ref{tab:bM} (i.e. $b_M^2 = 0.32$ and $b_M^3 = 0.3$).  It
is clear by eye that the data do not favor the higher bias.  Formally,
this model is disfavored at $1.9\sigma$.  Although there appears to be some tension between the fiducial $w(\theta)$ model and the measurements shown in Fig.~\ref{fig:wtheta} (in the opposite sense to that required for a higher mass bias), this tension is not statistically significant (note that the error bars in the different angular bins are highly correlated).  Specifically, the amplitude of the measured $w(\theta)$ differs from the fiducial model by only $1.4\sigma$ ($0.7\sigma$) in the lower (higher) stellar mass bin.  We thus conclude that the fiducial model is acceptable, and it is
unlikely that the evidence for $\alpha_{ub} > 0$ and $\alpha_{cb} > 0$ in the $y$-group cross-correlation function interpretation
is entirely due to unaccounted-for mass biases.  However, at the
current signal-to-noise level, it is possible that some of the
evidence can be attributed to larger mass biases than assumed in our
fiducial model.

\subsection{Thermal SZ Stacking on Locally Brightest Galaxies}
\label{sec:LBGs}

As described in \S\ref{sec:data}, \citetalias{PlanckLBG2013} used a
stacking analysis to measure the $y$ signal around a sample of ``locally
brightest galaxies'' selected from SDSS data.  To characterize the $y$
signal around these galaxies, \citetalias{PlanckLBG2013} relied on
both aperture photometry and a matched filter technique.  The inferred
$y$ measurements for individual halos were then stacked (i.e., averaged)
across halos of similar stellar mass.  Finally, a stellar mass--halo
mass relation was used to infer the relationship between halo mass and
tSZ flux.

The results in Fig.~\ref{fig:data} and our discussion of the halo
model fits make it clear that modeling the two-halo contribution is
essential in fitting the $Y$--$M$ relation. At halo masses below 
$\approx 10^{13}\,h^{-1}M_{\odot}$ the two-halo term dominates the signal at
all scales in the Planck measurement shown in Fig.~\ref{fig:data}.  Consequently, a significant
difficulty associated with the stacking approach used by
\citetalias{PlanckLBG2013} is that the measured $y$ signal around a
galaxy will be contaminated by the $y$ signal from nearby halos.  To
prevent such contamination, \citetalias{PlanckLBG2013} imposed an
isolation requirement on their halo catalog to remove galaxies with
nearby, bright neighbors (with some cost in the signal-to-noise), as
described in \S\ref{sec:data}.  The fiducial isolation region imposed
by \citetalias{PlanckLBG2013} is a cylinder oriented along the
line-of-sight with transverse radius $R_{\rm iso} = 1\,{\rm Mpc}$ and
length $v_{\rm iso} = c \Delta z = 1000\,{\rm km}/{\rm s}$.  LBGs are
defined to be galaxies brighter in $r$-band magnitude than all other
galaxies within this isolation region.

It is not obvious, however, that the isolation criterion imposed by
\citetalias{PlanckLBG2013} is sufficient to ensure that the measured
$y$ signal for a halo is uncontaminated by the signal from neighbors
(i.e., the two-halo term).  The $y$ signal for a massive nearby halo
can extend well beyond the 1 Mpc isolation radius.  As one considers
halos of lower masses, the isolation criterion becomes more suspect
since the amplitude of the $y$ profile falls rapidly with decreasing
mass and the two-halo contribution becomes significant (or even
dominant) at small radii (e.g., see the lowest mass bins in
Fig.~\ref{fig:data}).  Furthermore, the \citetalias{PlanckLBG2013}
isolation requirement considers a galaxy to be isolated if it is
brighter in $r$ than nearby galaxies within the exclusion volume.
However, the two-halo term around high-mass halos receives a dominant
contribution from {\it lower}-mass halos since these are much more
abundant (see, e.g., Fig.~11 of \citet{Vikram2017}).  Consequently, the
\citetalias{PlanckLBG2013} isolation requirement will not remove
the two-halo contribution for high-mass halos.

\begin{figure}
\centering
\includegraphics[width=0.7\textwidth]{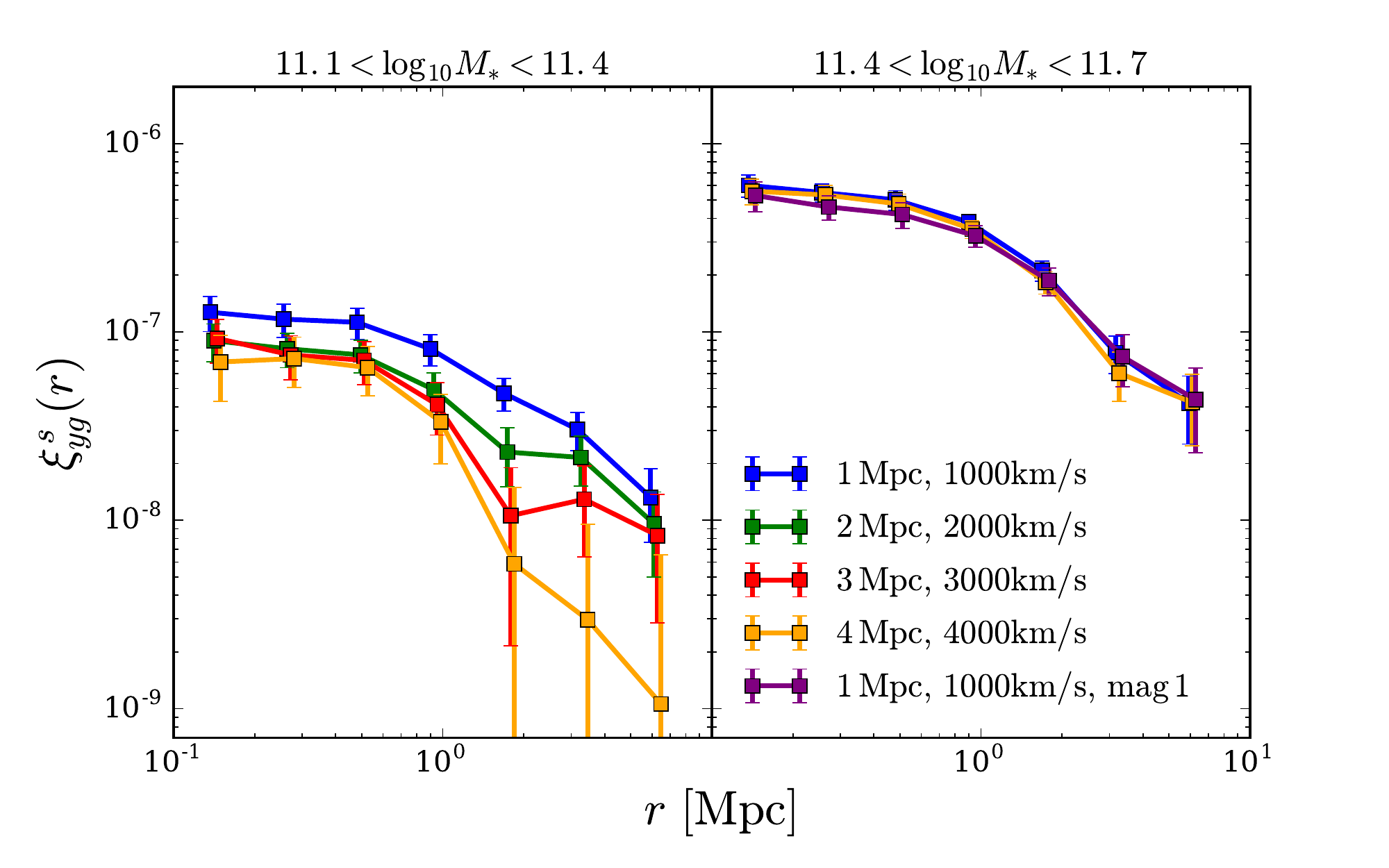}
\caption{\label{fig:LoBaG} The measured LBG-$y$ correlation for LBGs
  with stellar mass $11.1 < \log_{10} M_* < 11.4$ (left panel) and
  $11.4 < \log_{10} M_* < 11.7$ (right panel).  These correspond to halo
  masses of roughly $10^{12.5} \, M_{\odot} < M < 10^{14} \, M_{\odot}$
  and $10^{13.5} \, M_{\odot} < M < 10^{14.5} \, M_{\odot}$, respectively (see
  Fig.~B.1 of~\citetalias{PlanckLBG2013}).  LBGs are considered
  isolated if they are brighter in $r$ than all other galaxies within
  the isolation radius $R_{\rm iso}$ and within a redshift range
  $\Delta z = v_{\rm iso} /c$.  The different curves show results for
  different choices of $\left(R_{\rm iso}, v_{\rm iso} \right)$.  As
  the exclusion region is increased in size, the $y$ profile around
  LBGs in the lower mass bin decreases.  This suggests
  that the two-halo term makes a non-negligible contribution to the
  $y$ profiles around LBGs with the nominal exclusion choice,
  $\left(R_{\rm iso}, v_{\rm iso} \right) = \left( 1\,{\rm Mpc},
  1000\,{\rm km}/{\rm s} \right)$.  The purple curve in the right
  panel shows the effects of modifying the exclusion criterion so that
  an isolated galaxy must be more than one magnitude brighter in $r$
  than all other galaxies within the exclusion region. }
\end{figure}

Fig.~\ref{fig:LoBaG} demonstrates how the isolation criterion imposed
by \citetalias{PlanckLBG2013} may not be sufficient to ensure an
uncontaminated measurement of the $y$ signal from a low-mass halo.  We
plot the $y$-LBG correlation function for two stellar mass bins as the
isolation threshold is varied from $\left(R_{\rm iso}, v_{\rm iso}
\right) = \left( 1\,{\rm Mpc}, 1000\,{\rm km}/{\rm s} \right)$ to
$\left(R_{\rm iso}, v_{\rm iso} \right) = \left( 4\,{\rm Mpc},
4000\,{\rm km}/{\rm s} \right)$.  The left-hand panel of
Fig.~\ref{fig:LoBaG} makes it clear that when the isolation threshold
is increased to $\left( 2\,{\rm Mpc}, 2000\,{\rm km}/{\rm s} \right)$,
the LBG-$y$ correlation decreases at small scales for galaxies with
stellar mass $11.1 < \log_{10} M_{*} < 11.4$ (corresponding roughly to halo masses $M_{500} \approx 10^{12.5}$--$10^{14}$ $M_{\odot}$).  This suggests that
there is a residual two-halo contribution to the $y$-profile around
these ``isolated'' galaxies for the fiducial isolation criterion.  As
the isolation threshold is increased further, the $y$ profile at small
scales remains roughly constant, suggesting that most of the two-halo
term has been removed for isolation radii greater than $2\,{\rm Mpc}$.
We note that as the isolation criteria are varied, the mean stellar
mass of the LBG sample remains approximately constant, suggesting that
variation in the mean mass of the LBGs is not responsible for the
observed change in $y$ with varying isolation criterion.

The right-hand panel of Fig.~\ref{fig:LoBaG} shows that for galaxies
with higher stellar mass ($11.4 < \log_{10} M_* < 11.7$, corresponding roughly to halo masses $M_{500} \approx 10^{13.5}$--$10^{14.5}$ $M_{\odot}$), increasing the isolation radius has a
negligible impact on the $y$ profile.  This likely occurs because, in
the case of the more massive host halos in this LBG bin, the one-halo
term dominates over the two-halo contribution out to scales close to
that of the isolation radius. Further, we expect little if any change
to the $y$-group correlation function on scales much larger than the
isolation radius.  However, as pointed out above, for high-mass halos,
the two-halo term can receive a dominant contribution from
less-massive halos.  To explore this possibility, we re-generate the
LBG catalog modifying the isolation requirement such that galaxies are
considered isolated if they are the only galaxy {\it more than one
  magnitude} brighter than all other galaxies within the isolation
volume.  The LBG-$y$ correlation for this sample is shown as the
purple curve in the right panel of Fig.~\ref{fig:LoBaG}.  We find that
for this enhanced isolation criterion, the amplitude of the LBG-$y$
correlation function decreases, as expected if less
massive halos contribute significantly to the two-halo term for this
bin.

We note that \citetalias{PlanckLBG2013} also tested the effects of
varying the isolation criteria in their analysis, finding little
change in the inferred $Y$--$M$ relation, although they only considered
the $\left(R_{\rm iso}, v_{\rm iso} \right) = \left( 2\,{\rm Mpc},
2000\,{\rm km}/{\rm s} \right)$ case.\footnote{A detailed examination
  of Fig.~A.1 of~\citetalias{PlanckLBG2013} suggests that their inferred $Y_{500}$ values do
  indeed decrease somewhat when applying the stricter isolation
  criteria, for the halos corresponding to the stellar mass bin in the
  left panel of Fig.~\ref{fig:LoBaG} (corresponding roughly to
  $M_{500} \approx 10^{12.5}$--$10^{14}$ $M_{\odot}$). However, the
  changes are not statistically significant.}  The difference between
our findings and those of \citetalias{PlanckLBG2013} may be partially
due to the fact that they characterized the $y$ signal using matched filters and aperture
photometry.  These techniques will act to reduce
contamination from a two-halo term that is fairly flat with radius.  In Appendix~\ref{app:MF}, we consider this point in some detail.
However, as pointed out by \citet{LeBrun2015}, the matched filter
approach used by \citetalias{PlanckLBG2013} can be quite sensitive to
assumptions about the profile shape below the angular resolution
limit of Planck.  We emphasize that the measurement presented here is
intended only to demonstrate that the isolation criterion imposed by
\citetalias{PlanckLBG2013} is likely insufficient to remove all of the
two-halo contribution to the $y$ profile for low-mass halos, even at
small radii.

Finally, we emphasize that imposing an isolation requirement in an
attempt to suppress the two-halo term is non-optimal in terms of
signal-to-noise of the measurements.  By definition, an isolated
catalog will contain fewer objects and will therefore throw out
potentially useful signal.  The approach of modeling the two-halo
term, on the other hand, can exploit all of the available information
in the full (non-isolated) catalog.

\section{Interpretation and Outlook}
\label{sec:interp}

The results presented in \S\ref{sec:analysis} indicate that the two-halo term is non-negligible for stacked tSZ measurements around halos below the scale of massive galaxy clusters.  Modeling this term and accounting for it when fitting data to theoretical predictions could significantly impact conclusions about the $Y$--$M$ relation, and thus the influence of AGN and supernova feedback on the hot gas distribution.  Fitting various theoretical models (accounting for both the one-halo and two-halo terms) to the $y$-group cross-correlation function, we find moderate ($\approx 2\sigma$) evidence for a suppression in the thermal gas pressure in low-mass ($M \lesssim 10^{14} \, M_{\odot}$) systems, relative to the prediction of near-self-similar (break-free) models.  The inferred $Y$--$M$ relation in the context of the break models (\textit{UB} and \textit{CB}) is qualitatively consistent with predictions from cosmological simulations incorporating energetic feedback~(e.g.,~\cite{LeBrun2015,LeBrun2017}).  Note that the evidence of deviation from self-similar mass dependence presented here can be explained by multiple effects (or combinations thereof): the gas could simply be depleted from low-mass halos, or the gas could be present but at unexpectedly low temperatures in such halos, with a correspondingly large amount of non-thermal pressure support required to counteract gravity (i.e., large hydrostatic mass bias, due to turbulent motions, magnetic fields, etc.).  All of these effects act to reduce the observed tSZ signal.  At the current level of precision, it is beyond the scope of our analysis to make a statement about the possible origin of the deficit, which would furthermore require non-tSZ data.  However, it is worth noting that mass-dependent hydrostatic mass bias could be a significant cause for concern in tSZ- and X-ray-based cluster cosmology analyses, if its effects were non-negligible at cluster mass scales.

A pure power-law $Y$--$M$ relation also fits the data well, with a slope consistent with the self-similar value --- however, the power-law fit is driven by the measurements for the highest-mass halos in the sample, where (near-)self-similarity is already known to be a good description.  Our fits to the break models show that evidence for deviations from self-similarity may be present in existing data, but can be obscured by fits to simple model parameterizations, such as a pure power-law.  There are many additional model variations that can be considered beyond the break models considered here, but improved data will be necessary to constrain models with more than one free parameter.

We also test for the impact of the uncertainty in the group--halo mass relation on our results via the relationship between halo mass and the bias inferred from the group clustering at large scales. The $w(\theta)$ results in Fig.~\ref{fig:wtheta} suggest that a bias in the halo mass estimates is not solely responsible for the evidence we find for non-self-similarity in the tSZ data.  At the current level of precision, such degeneracies are difficult to break: higher signal-to-noise measurements at higher resolution will be needed to definitively determine the behavior of the pressure--mass relation below cluster mass scales.  A promising avenue for future work would be a joint analysis of the $y$-group cross-correlation, group auto-correlation, and group-lensing cross-correlation functions, which would further break degeneracies in the current modeling.

Fortunately, there are excellent near-future prospects for constraining the distribution of hot gas in low-mass halos using an analysis similar to the approach developed here (see also~\citet{Vikram2017}). In particular, the current measurements are in large part limited by the relatively coarse Planck beam. This is insufficient to resolve the one-halo term in low-mass groups, which may reveal the sharpest departures from self-similarity.  It will therefore be interesting to pursue stacking measurements around low-mass groups using higher-resolution CMB data from the Atacama Cosmology Telescope (ACT) and South Pole Telescope (SPT), with beams of FWHM $\approx 1$ arcmin compared to the FWHM = 10 arcmin resolution of the Planck $y$ maps. In addition, future high-resolution CMB experiments such as Simons Observatory\footnote{\url{http://www.simonsobservatory.org}}, CCAT-prime\footnote{\url{http://www.ccatobservatory.org/}}, and CMB-S4\footnote{\url{http://www.cmb-s4.org}} will contain additional frequency channels, which will aid in separating the tSZ signal from the cosmic infrared background and other potential contaminants.  The higher sensitivity of these measurements will also allow more freedom in the modeling, such as simultaneously fitting the break mass and power-law slope in the \textit{UB} and \textit{CB} models considered here (rather than fixing the break mass to $M_0 = 10^{14} \, M_{\odot}$).  Likewise, simultaneously fitting additional tSZ statistics, such as the tSZ auto-power spectrum~(e.g.,~\cite{Komatsu2002,Hill-Pajer2013,Dolag2016}), tSZ -- lensing cross-correlations~\cite{Hill-Spergel2014,vWHM2014,Hojjati2016,BHM2015}, $\langle y \rangle$~(e.g.,~\cite{Hill2015,Dolag2016}), and higher-order statistics~(e.g.,~\cite{Wilson2012,Hill-Sherwin2013,Bhattacharya2012,Crawford2014,Hill2014}), will provide further constraints on feedback models, with the additional benefit that direct halo mass estimates will not be required.  Joint analyses with kinematic SZ measurements will also be informative in this regard~(e.g.,~\cite{Battaglia2017}), using methodology similar to that developed in this paper.

\begin{figure}
\begin{center}
\includegraphics[trim=0cm 6cm 0cm 3cm, width=0.6\textwidth]{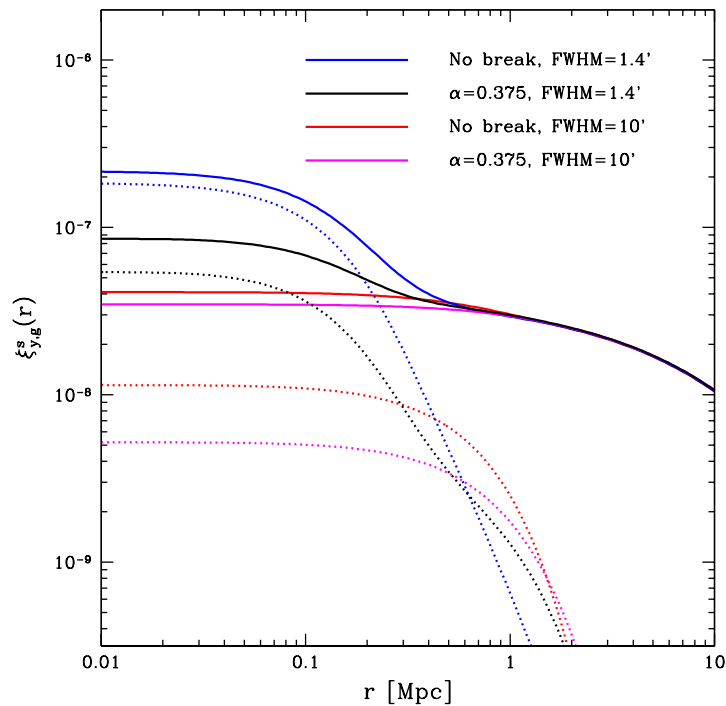}
\end{center}
\caption{\label{fig:prospects} Future prospects for constraining the distribution of hot gas in low-mass groups. Here we contrast the predicted $y$-group cross-correlation for our $12 < \log_{10} M/(h^{-1} M_\odot) < 13$ mass bin for the fiducial Battaglia (break-free) model and an example \textit{CB} model, with an assumed sub-break power-law of $\alpha_{cb}=0.375$.  We show results for Planck angular resolution (FWHM=10 arcminutes) and ACT/SPT resolution (FWHM=1.4 arcminutes).  The solid curves show the total $y$-group cross-correlations, while the dotted curves show the one-halo terms. The blue and red curves show the Battaglia pressure profile prediction at higher and lower angular resolution, respectively, while the black and magenta curves are the \textit{CB} models at higher and lower resolution, respectively. The higher angular resolution of ACT/SPT helps to partly resolve the one-halo term in these low-mass systems. This should allow sharper tests of self-similarity in the near future. Note that the models here ignore any miscentering effects, assuming this may be well-calibrated using lensing measurements.}
\end{figure}

In order to illustrate the potential utility of these future measurements, Fig.~\ref{fig:prospects} shows the fiducial Battaglia model and an example \textit{CB} model at Planck and ACT/SPT angular resolutions in our  $12 < \log_{10} M/(h^{-1} M_\odot) < 13$ mass bin.\footnote{Note that the relevant resolution here is that of the component-separated maps. This should be comparable to, but perhaps slightly worse than, the value of FWHM=1.4 arcminutes adopted here for ACT/SPT.} \citet{Vikram2017} also considered the benefits of improved angular resolution in probing the hot gas in low-mass halos, but here we further explicitly illustrate the power of these measurements for studying departures from self-similarity.  At Planck resolution (FWHM=10 arcminutes), the difference between the example models is extremely subtle: in this case, even the inner radial bins are dominated by the two-halo term from massive neighbors. At the resolution of ACT/SPT, however, the inner radial bins become one-halo-dominated even for these low-mass halos and it should therefore be possible to distinguish the example models at high statistical significance. The results in Fig.~\ref{fig:prospects} motivate the need for $\sim$arcminute-scale resolution in upcoming CMB experiments.  In conjunction with the improved CMB data, it will be important to improve the calibration of the average group mass--halo mass relation, its scatter, and miscentering errors. These steps should be possible with galaxy-galaxy lensing and clustering measurements, performed in tandem with additional modeling efforts.  It will also be important to cross-check several simplifying assumptions made in the halo model, such as linear biasing and the neglect of halo exclusion, with numerical simulations.  Lastly, we note that there have been few attempts since~\citet{Yang2007} to build a halo catalog spanning the full range from galaxy- to cluster-scale masses (see, however, the recent catalog of~\citet{Lim2017}), and additional work to understand the different systematics that could arise over a such a large mass range may be necessary.

Overall, significant developments in the study of the cosmic distribution of hot gas should be forthcoming in the next few years.  Deeper optical observations from the Dark Energy Survey~\cite{DES2005}, Hyper Suprime-Cam Survey~\cite{HSC2017}, Kilo Degree Survey~\cite{KiDS2013}, and Large Synoptic Survey Telescope~\cite{LSST2009}, as well as application to quasar samples, will allow these probes to be extended to much higher redshifts.  Given the expected improvement in experimental sensitivity and resolution, precise modeling of the signal is important.  In this context, we have shown that the two-halo term in tSZ cross-correlation or stacking measurements cannot be neglected (except for perhaps the most massive galaxy clusters) and should be included in future analyses.

\begin{acknowledgments}
We are very grateful to Vinu Vikram for collaborative work on an
earlier study and several helpful discussions.  We thank James
Bartlett, Nick Battaglia, Mike Jarvis, Niall MacCrann, Ravi Sheth, Wenting Wang, and
Simon White for useful conversations.  We are also grateful to the anonymous referee for feedback that improved the manuscript.  JCH thanks the University of
Pennsylvania Department of Physics and Astronomy for hospitality
during the course of this research.  This work was partially supported
by a Junior Fellow award from the Simons Foundation to JCH.  EB and BJ
are partially supported by the US Department of Energy grant
DE-SC0007901 and funds from the University of Pennsylvania. JPG is supported by the National Science Foundation Graduate Research Fellowship under Grant No. DGE 1148900.
\end{acknowledgments}

\begin{appendix}
\section{Constraints on the Normalization of the Pressure--Mass Relation}
\label{app:P0}

\begin{table}[ht]
\begin{center}
  MILCA \vspace{2pt} \\
  \begin{tabular}{| c | c | c |}
    \hline 
     Model & Parameter [prior range] & Marginalized constraint \\
     \hline \hline
    {\it PL} & $\alpha_{pl}$ [-1, 1]  & $-0.003 \pm 0.06$ \\ 
                & $P_0^{pl}$ [0.2, 1.8] & $0.93 \pm 0.06$ \\ \hline
    {\it UB} & $\alpha_{ub}$ [-1, 1.25] & $0.27_{-0.22}^{+0.23}$ \\
                & $P_0^{ub}$ [0.2, 1.8] & $0.97 \pm 0.05$ \\ \hline
    {\it CB} & $\alpha_{cb}$ [0, 2] & $0.58 \pm 0.34$ \\
                & $P_0^{cb}$ [0.2, 1.8] & $0.97_{-0.04}^{+0.05}$ \\
    \hline
  \end{tabular}
  \\ \vspace{12pt} NILC \vspace{2pt} \\
  \begin{tabular}{| c | c | c |}
    \hline 
     Model & Parameter [prior range] & Marginalized constraint \\
     \hline \hline
    {\it PL} & $\alpha_{pl}$ [-1, 1]  & $-0.005 \pm 0.06$ \\ 
                & $P_0^{pl}$ [0.2, 1.8] & $0.88 \pm 0.06$ \\ \hline
    {\it UB} & $\alpha_{ub}$ [-1, 1.25] & $0.29_{-0.21}^{+0.22}$ \\
                & $P_0^{ub}$ [0.2, 1.8] & $0.92 \pm 0.05$ \\ \hline
    {\it CB} & $\alpha_{cb}$ [0, 2] & $0.35 \pm 0.23$ \\
                & $P_0^{cb}$ [0.2, 1.8] & $0.91\pm 0.04$ \\
    \hline
  \end{tabular}
  \caption{Constraints on the mass dependence and overall normalization of the electron pressure profile for various theoretical models (see \S\ref{sec:halo} and Appendix~\ref{app:P0} for model and parameter definitions).  The fiducial Battaglia model in all cases corresponds to $\alpha_{pl} = \alpha_{ub} = \alpha_{cb} = 0$ and $P_0^{pl} = P_0^{ub} = P_0^{cb} = 1$.  The final column presents constraints on $\alpha$ ($P_0$) after marginalizing over $P_0$ ($\alpha$) and marginalizing over the correctly-centered fraction of halos in each of the five mass bins, with an uninformative prior on the centered fraction $p_c^I \in [0,1]$ for all bins.  The tabulated values are the mean and $68\%$ C.L. intervals computed from the one-dimensional marginalized posteriors.}
  \label{tab:alphaP0}
\end{center}
\end{table}

Here we consider a further extension of the models presented in \S\ref{sec:halo}, in which the overall normalization of the pressure--mass relation ($P_0$) is also allowed to vary.  The primary focus of our analysis above was to use the large lever arm in mass of the \citet{Yang2007} group sample to probe the mass dependence of the tSZ signal and test whether it follows self-similar predictions.  However, the $y$-group cross-correlation is also sensitive to the overall normalization of the pressure--mass relation.  Due to complications discussed in \S\ref{sec:halo} (e.g., miscentering, the group mass -- halo mass relation, etc.), this method of calibrating the overall normalization is less direct and precise than using gravitational lensing observations of tSZ clusters.  Thus, although we present quantitative results here (in the context of the underlying Battaglia model), we caution against a face-value interpretation of the results.

In these extended models, the equations describing the pressure profile behavior take on the following forms (see \S\ref{sec:halo} for futher background):
\begin{itemize}
\item {\bf {\it PL} Model}:
\beqa
P_e(r|M,z) \rightarrow P_e(r|M,z) \, P_0^{pl} \left( \frac{M}{M_0} \right)^{\alpha_{pl}} \,.
\label{eq:PLP0}
\eeqa
\item {\bf {\it UB} Model}:
\beqa
P_e(r|M,z) \rightarrow
\begin{cases}
P_e(r|M,z)\, P_0^{ub} \,, & M \geq M_0 \\
P_e(r|M,z)\, P_0^{ub} \left( \frac{M}{M_0} \right)^{\alpha_{ub}} \,, & M < M_0 \,.
\end{cases}
\label{eq:UB}
\eeqa
\item {\bf {\it CB} Model}:
\beqa
P_e(r|M,z) \rightarrow
\begin{cases}
P_e(r|M,z)\, P_0^{cb} \,, & M \geq M_0 \\
\left( P_e(r|M,z) \left( \frac{M}{M_0} \right)^{\alpha_{cb}} + A(\alpha_{cb}|M,z) \, e^{\frac{-r^2}{2(2r_{\rm vir})^2}} \right) P_0^{cb} \,, & M < M_0 \,.
\end{cases}
\label{eq:CB}
\eeqa
\end{itemize}

The marginalized constraints on the amplitude and mass-dependence parameters in these models are given in Table~\ref{tab:alphaP0}.  In all cases, the amplitude constraints are consistent with the fiducial model ($P_0 = 1$), although the central values tend to lie slightly below unity, implying a lower normalization than that in the Battaglia model.  Interestingly, the statistical uncertainty ($\approx 5$\%) is competitive with weak-lensing calibrations of the normalization of the $Y$--$M$ relation at cluster mass scales.  However, as emphasized above, we have not considered and marginalized over all relevant systematic effects here.

After marginalizing over $P_0$, the error bars on $\alpha$ do not dramatically increase, if at all.  For the {\it PL} model, the error bars on $\alpha_{pl}$ increase by $\approx 50$\% compared to those in Table~\ref{tab:alpha}, where $P_0^{pl}$ was fixed to unity.  However, the central value of $\alpha_{pl}$ moves toward $\alpha_{pl} \approx 0$, which brings the best-fit {\it PL} model predictions somewhat closer to those for the {\it UB} and {\it CB} models in the low-mass bins.  Effectively, the free amplitude absorbs some of the statistical constraining power from the high-mass bins, allowing the $\alpha_{pl}$ value to move closer to the values preferred by the lower-mass bins, which are closer to $\alpha_{pl} \gtrsim 0$.  The resulting $\alpha_{pl}$ values are consistent with the Battaglia model or the self-similar prediction.

For the {\it UB} and {\it CB} models, the error bars on $\alpha$ do not increase compared to those in Table~\ref{tab:alpha}.  This is simply because the highest two mass bins constrain $P_0$ in these models, while the lower mass bins constrain $\alpha$, effectively independently of one another.  However, the central values of $\alpha_{ub}$ and $\alpha_{cb}$ move toward zero compared to those in Table~\ref{tab:alpha} (i.e., the evidence of a departure from self-similar mass-dependence weakens somewhat).  This is because the high-mass bins' slight preference for $P_0 < 1$ implies that less deviation from $\alpha = 0$ is needed in order to fit the signal in the lower-mass bins.  This result is seen for both the {\it UB} and {\it CB} models.  Lastly, we note that the surprisingly small error bar on $\alpha_{cb}$ for the NILC case is due to the influence of the prior range for this parameter on the posterior (i.e., the hard cutoff at $\alpha_{cb} = 0$).

Overall, the analysis presented here serves as a consistency check that our data are not fully explained by self-similarity with a lower overall normalization (which could be interpreted as a fixed, mass-independent amplitude of non-thermal pressure) in the context of the {\it UB} and {\it CB} models.  However, allowing the normalization to vary does slightly weaken the evidence of a departure from self-similar mass dependence in these models.

\section{Estimate of Matched Filter Suppression of the Two-Halo Term}
\label{app:MF}
Here we investigate the efficacy with which tSZ matched filters --- applied to Planck Compton-$y$ maps around galaxy groups --- suppress contributions from the two-halo term. We consider applying a filter to a 2D $y$-map centered on a galaxy group, such that the data $d({\pmb \theta})$ is the sum of some known pressure profile, with a projected and beam-smoothed Compton-$y$ profile given by $y_s({\pmb \theta})$, and ``noise", $N({\pmb \theta})$:
\beqa
d({\pmb \theta}) = y_s({\pmb \theta}) + N({\pmb \theta}).
\label{eq:data}
\eeqa
In this equation, the noise term is understood to include everything in the map that is not part of the $y$-profile associated with the galaxy group of interest. This includes instrumental noise, $y$ fluctuations from other systems, and residual foregrounds. Note that the noise is assumed to have zero mean, which is violated by the {\em correlated} ``noise'' from the two-halo term. This effect has not been considered explicitly in previous literature related to tSZ matched filters, and it may produce a bias.  On the other hand, the filters as usually constructed may not, in practice,  pass much of the two-halo term.  Our goal here is to assess this quantitatively, particularly with regard to the analysis of~\citetalias{PlanckLBG2013}.\footnote{We are grateful to Simon White and the anonymous referee for suggesting this calculation.}

Before proceeding, it is also important to note that \citetalias{PlanckLBG2013} applies matched filters directly to the multi-frequency Planck temperature maps and not to the $y$-map itself, which is smoothed to the resolution of the coarsest HFI channel. Furthermore, the isolation criteria applied in constructing the LBG sample used in the~\citetalias{PlanckLBG2013} analysis will lessen the impact of the two-halo term to some extent, as investigated in \S\ref{sec:LBGs}. Using the higher resolution multi-frequency maps and applying the isolation criteria should both act to reduce the two-halo contamination studied here. We briefly consider this point at the end of this appendix.

Under the standard assumptions, the optimal matched filter is given by (e.g.,~\cite{Haehnelt1996,Herranz2002}):
\beqa
F({\sc l}) = A \frac{y({\sc l}) B_{\sc l}} {B^2_{\sc l} C^{\rm sky}_{\sc l} + N_{\sc l}} \,,
\label{eq:match}
\eeqa
where $y({\sc l})$ is the 2D Fourier transform of the assumed (projected) pressure profile, $B_{\sc l}$ gives the Planck beam profile in Fourier space, $C^{\rm sky}_{\sc l}$ describes the angular power spectrum of emission fluctuations in the Compton-$y$ map (including residual foregrounds), and $N_{\sc l}$ is the instrumental noise power spectrum. In the calculations that follow (apart from estimating the power spectra of the Planck $y$-maps), we adopt the flat-sky approximation. In constructing tSZ matched filters, we follow~\citet{PlanckLBG2013} and use the~\citet{Arnaud2010} pressure profile, truncated at $6 r_{500}$, to calculate $y({\sc l})$ (according to, e.g., Eq. 2 of \cite{Komatsu2002}).  We assume a Gaussian beam, $B_{\sc l} = {\rm exp}[-{\sc l}({\sc l}+1) \sigma^2/2]$ where $\sigma = {\rm FWHM}/\sqrt{8 {\rm ln(2)}}$, with ${\rm FWHM} = 10$ arcmin, as appropriate for the Planck Compton-$y$ maps. The matched filter downweights modes where the noise term, $B^2_{\sc l} C^{\rm sky}_{\sc l} + N_{\sc l}$, is large compared to the template $y$-profile, $A y({\sc l}) B_{\sc l}$. In practice, we measure the auto-power spectrum of the Planck Compton-$y$ maps (including the noise bias), and deconvolve the beam such that the measured power spectrum is $C^{\rm tot}_{\sc l} = C^{\rm sky}_{\sc l} + N_{\sc l}/B^2_{\sc l}$. The filter is then constructed from the total, beam-deconvolved power spectrum, $C^{\rm tot}_{\sc l}$, as $F({\sc l}) = A y({\sc l})/(B_{\sc l} C^{\rm tot}_{\sc l})$.

The normalization constant, $A$, may be set to preserve the central, beam-smoothed value of the $y$-profile in configuration space after applying the matched filter, provided the observed pressure profile indeed follows the~\citet{Arnaud2010} form (see~\citet{LeBrun2015} for a discussion of biases that arise when this assumption is incorrect). This normalization requirement sets:
\beqa
A =  \left[ \int \frac{{\sc l} d{\sc l}}{2 \pi} \frac{y^2({\sc l}) B^2_{\sc l}}{B^2_{\sc l} C^{\rm sky}_{\sc l} + N_{\sc l}}\right]^{-1} \int \frac{{\sc l} d{\sc l}}{2 \pi} y({\sc l}) B_{\sc l} \,.
\label{eq:filt_norm}
\eeqa
One can also introduce an overall multiplicative constant in front of the template $y$-profile, varying this constant and perhaps the form of the profile itself to maximize the signal-to-noise ratio.  Here we confine our attention to the impact of matched-filtering on the two-halo term.

We construct the filter by measuring power spectra directly from the Planck $y$-maps, comparing NILC and MILCA to test sensitivity to the tSZ reconstruction algorithm. Before measuring the power spectra, bright regions in the Planck 857 GHz map are masked (along with a standard Planck point source mask) in order to remove Galactic dust emission, and the resulting mask is apodized using a Gaussian taper with FWHM = 15 arcmin.  We consider the power spectra measured for two different masked sky fractions, $f_{\rm sky}=0.3$ and $f_{\rm sky}=0.5$, in which the brightest $70\%$ and $50\%$ of the 857 GHz pixels are masked, respectively. The mask, beam, and pixel window functions are deconvolved using a MASTER-based code~\cite{MASTER2002}.  Our main results are insensitive to the choice of sky fraction. 

\begin{figure}
\begin{center}
\includegraphics[width=0.49\textwidth]{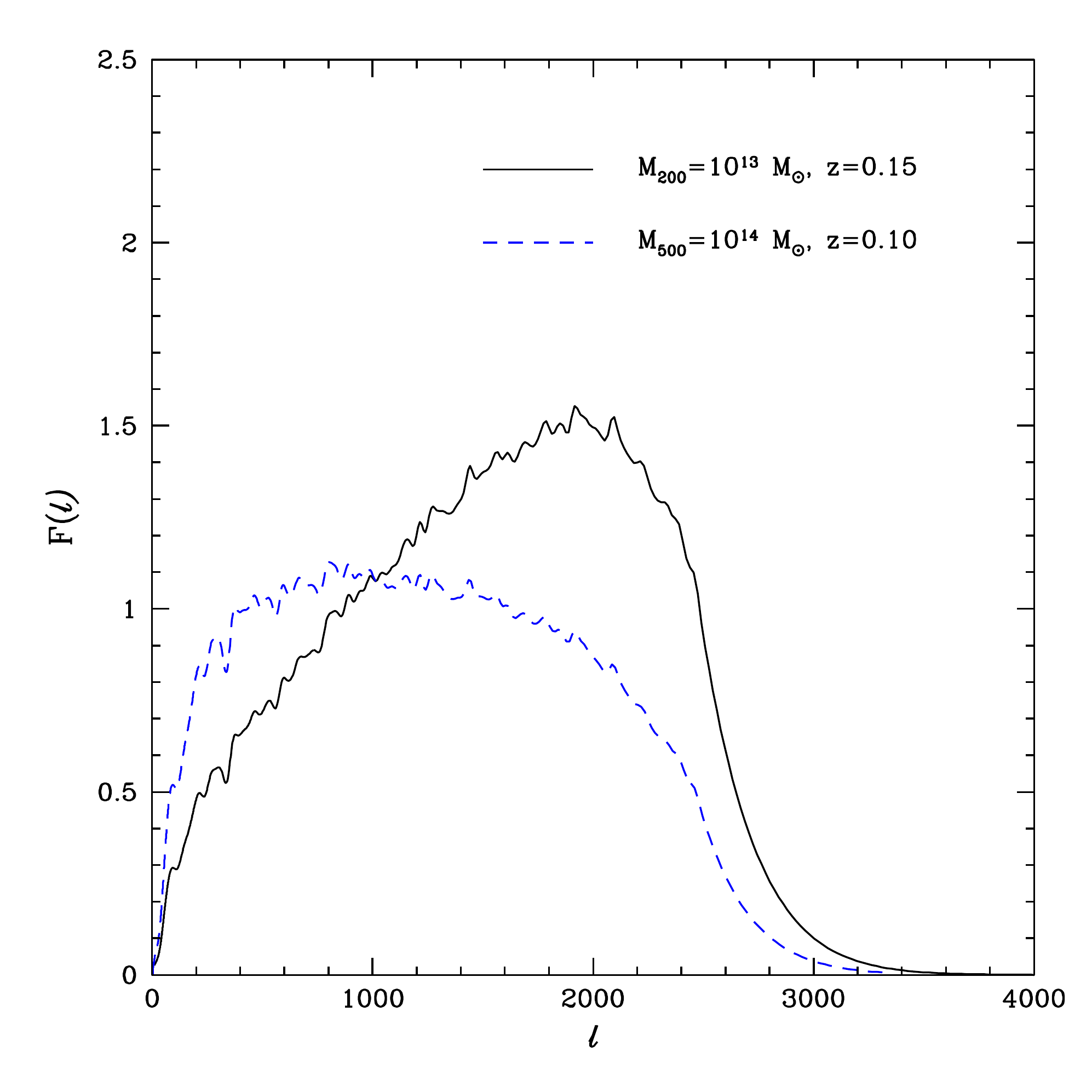}
\includegraphics[width=0.49\textwidth]{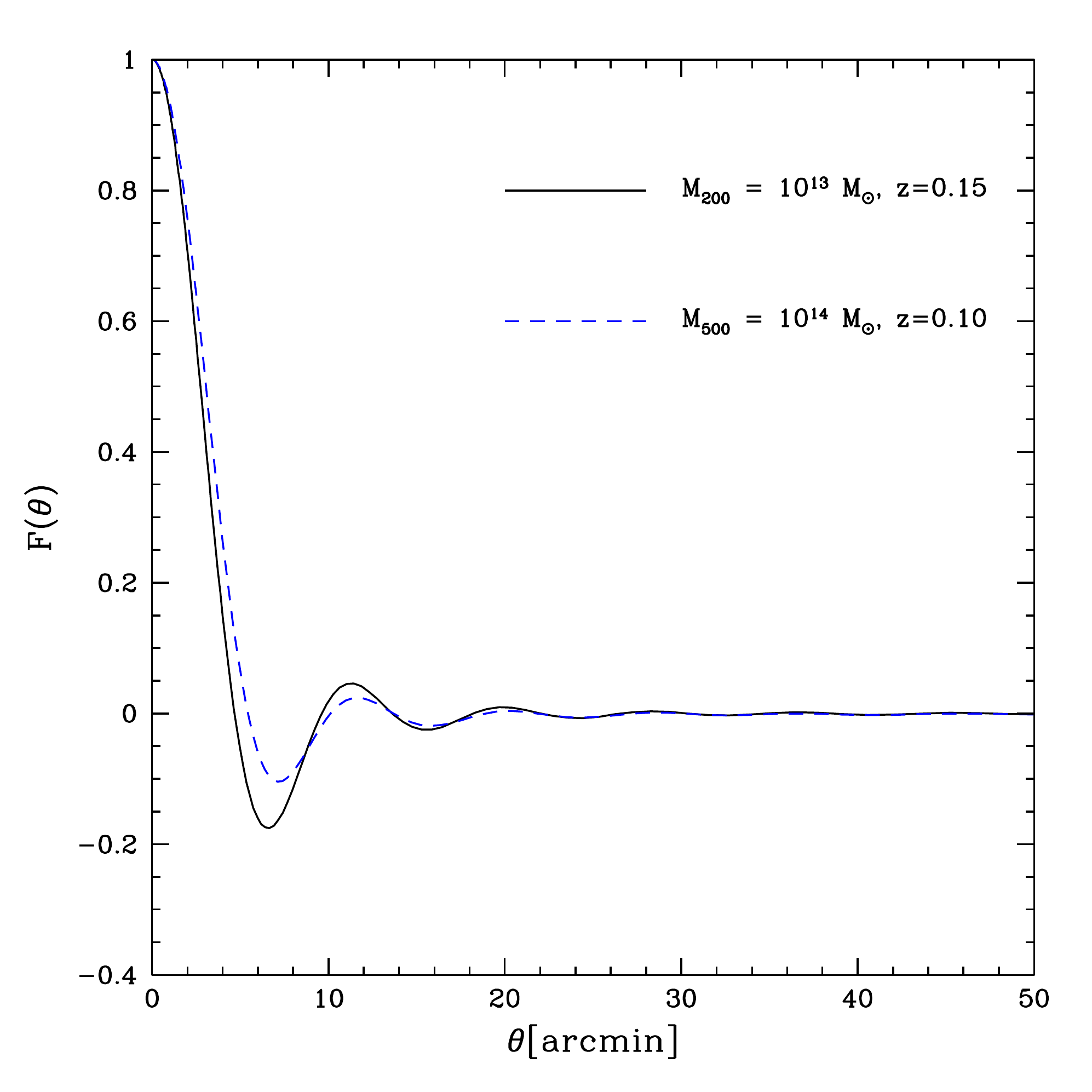}
\caption{Matched filters in Fourier space (left panel) and configuration space (right panel) for two example masses/redshifts, as labeled.  In the right panel, the filters are normalized to unity at $\theta=0$.}
\label{fig:filter_examples}
\end{center}
\end{figure}

Two examples of the resulting matched filters (Eq.~\ref{eq:match}) are shown in Fig.~\ref{fig:filter_examples}, both in Fourier space (left panel) and configuration space (right panel).  In the massive, lower-redshift case more large-scale (low-${\sc l}$) power is passed through the filter, as expected. In general, the optimal matched filters appear to pass some amount of low-${\sc l}$ power, which can thus include contributions from the two-halo term.

We can then calculate the effect of the matched filters on various models for the one- and two-halo terms. In practice, we carry out these calculations in Fourier space (where the configuration-space convolution is a simple product), and then Fourier transform back into configuration space. Denoting the $y$-halo cross-spectrum as $C_{x, \sc l}$, the filtered $y$-halo correlation function for a halo of mass $M$ and redshift $z$ at separation $\theta$ is:
\beqa
\xi^F_{y,g}(\theta|M,z) = \int \frac{{\sc l} d{\sc l}}{2 \pi} J_0({\sc l} \theta) C_{x, \sc l} B_{\sc l} F({\sc l}).
\label{eq:yg_filt}
\eeqa
Note that the matched filter acts on the Planck beam-smoothed field, hence two separate filters ($F({\sc l})$ and $B_{\sc l}$) are applied here to the intrinsic (i.e., unsmoothed) cross-spectrum.

\begin{figure} 
\begin{center}
\includegraphics[width=0.55\textwidth, trim={0 5cm 0 2.5cm}]{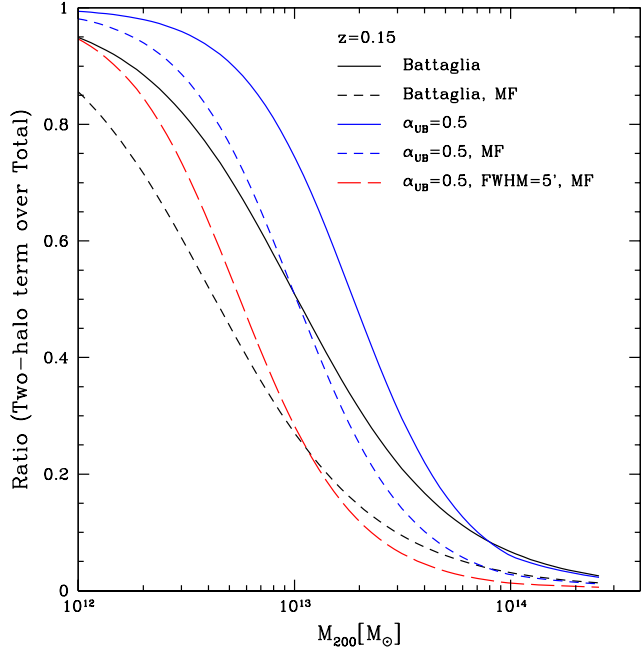}
\caption{Fractional contamination due to the two-halo term as a function of halo mass at $z=0.15$, with and without applying the matched filter for different pressure profile models. Each curve shows the fraction of the total $y$-halo cross-correlation signal contributed by the two-halo term for an inner radial bin (with $\theta \ll$ FWHM of the Planck beam). The black curves assume the fiducial Battaglia pressure profile, while the blue and red curves adopt the {\it UB} model with $\alpha_{ub}=0.5$ (see \S\ref{sec:halo}). The solid curves are smoothed only by the Planck $y$-map beam (FWHM = 10 arcmin), while the dashed curves show the effect of applying (in addition) the matched filter.  The long-dashed, red curve shows a modified matched-filter calculation in which the beam is assumed to have FWHM = 5 arcmin resolution (so as to capture some of the improvement associated with using the highest-resolution Planck HFI channels), but the filter itself is left unchanged (see Eq.~\ref{eq:yg_filt}).  Our primary conclusion is that although the matched filter suppresses the two-halo contamination, it is not entirely removed.}
\label{fig:ratio}
\end{center}
\end{figure}

As a first characterization of the impact of the filter on each of the one and two-halo terms, we consider the $\theta \rightarrow 0$ limit. The fractional contamination from the two-halo term depends strongly on the halo mass $M$, and somewhat on the underlying pressure profile model. To explore this, we vary $M_{200}$ from $10^{12} \, M_\odot$ to $10^{14.5} \, M_\odot$ while fixing $z=0.15$. Further, we consider models in which the pressure profile follows either the fiducial Battaglia fitting formula~\cite{Battaglia2012} or the {\it UB} model with $\alpha_{ub}=0.5$.  Because the pressure of low-mass halos is suppressed in the {\it UB} model, the two-halo term makes a larger fractional contribution than in the Battaglia model . The results of these calculations are shown in Fig.~\ref{fig:ratio}. Applying the matched filter suppresses the two-halo term by almost a factor of $\approx 3$, but it still comprises $\approx 50\%$ of the total signal (i.e., it is comparable to the one-halo contribution) in the $\alpha_{ub}=0.5$ model and is $\approx 25\%$ of the total signal in the Battaglia model, for a representative mass of $M_{200} = 10^{13} M_\odot$. At lower masses, the fractional contribution of the two-halo term is, of course, larger.\footnote{Note that at very high masses, near $M_{200} \approx 10^{14} M_\odot$, the fractional contribution of the two-halo term is {\em reduced} in the {\it UB} model. This is because the two-halo term is slightly suppressed in this model, while the one-halo term at high mass is preserved.}

In the actual~\citetalias{PlanckLBG2013} analysis, a multifrequency matched filter was adopted, for which the ``effective'' resolution is somewhat better than the FWHM = 10 arcmin resolution of the all-sky Planck $y$-maps, as the highest-resolution Planck HFI channels have FWHM $\approx$ 5 arcmin.  Thus, our matched-filter results are not exactly analogous to those in~\citetalias{PlanckLBG2013}.  However, constructing a new $y$-map with FWHM = 5 arcmin resolution, measuring its noise properties, and constructing the associated matched filter is beyond the scope of our analysis.  Instead, to capture some of the improvement associated with the higher effective resolution of the multifrequency matched filter approach used in~\citetalias{PlanckLBG2013}, we re-compute Eq.~\ref{eq:yg_filt} for a theoretical $y$-profile convolved with a FWHM = 5 arcmin beam, rather than a FWHM = 10 arcmin beam, while leaving the filter unchanged.  This is optimistic in that~\citetalias{PlanckLBG2013} use a 3-band multifrequency matched filter with bands at 100, 143, 217 GHz with FWHM = 9.68, 7.30, 5.02 arcmin, respectively: the ``effective'' beam is therefore coarser than the 5 arcmin value we adopt here.  But since we adopt a coarser-resolution filter $F(\ell)$, we neglect the additional improvement that would come from optimally modifying $F(\ell)$ for the higher-resolution channels.  The result of this calculation for the {\it UB} model is the red, long-dashed curve in Fig.~\ref{fig:ratio}.   As expected, this further reduces the fractional importance of the two-halo term. The only effect missing here is that there should be a further suppression from adjusting the matched filter for the finer beam.  We leave a full treatment of this issue (also including the isolation criteria for the~\citetalias{PlanckLBG2013} analysis) to future work.  Our primary conclusion remains unchanged: non-negligible two-halo contamination is still present, particularly at low masses.

\begin{figure}
\begin{center}
\includegraphics[width=0.49\textwidth]{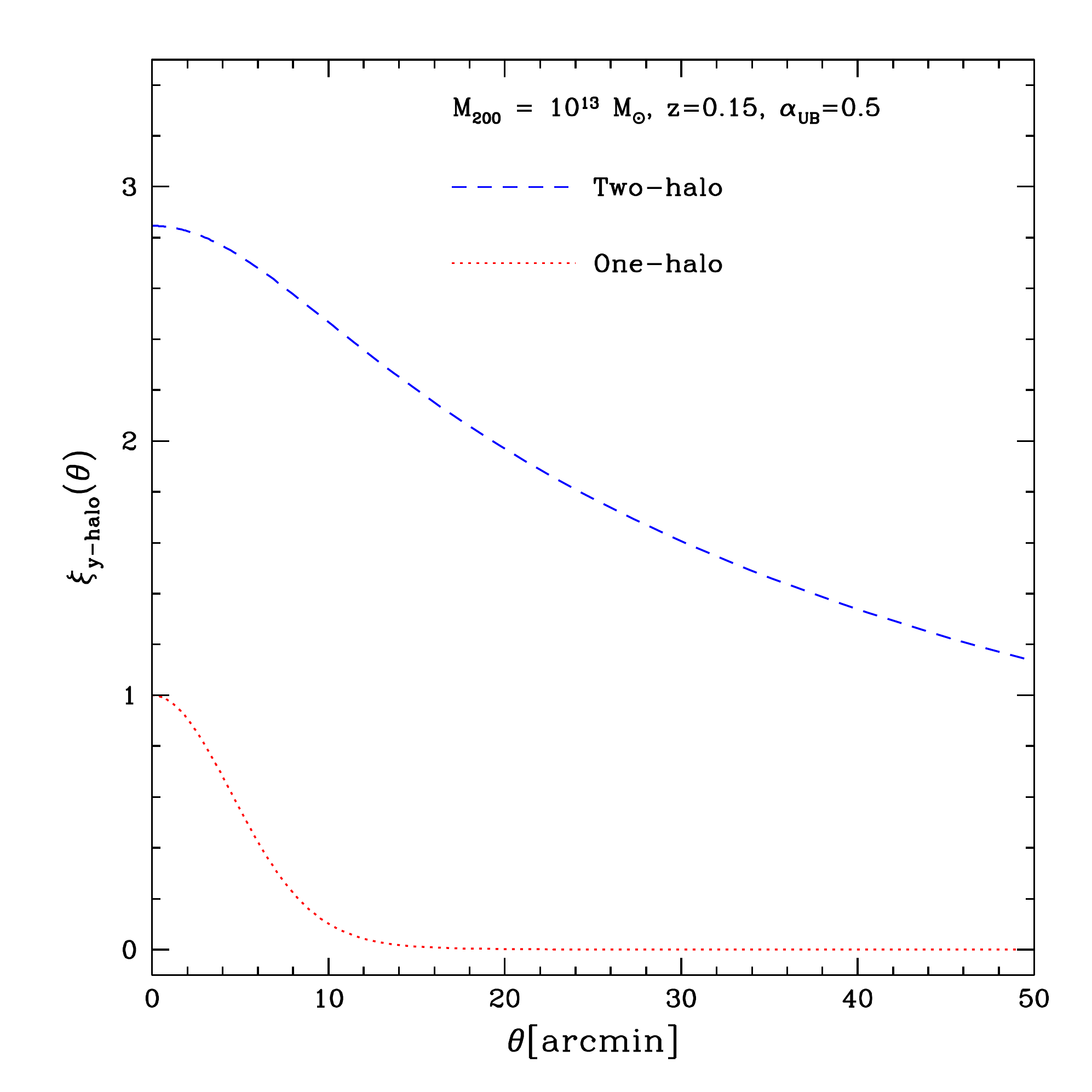}
\includegraphics[width=0.49\textwidth]{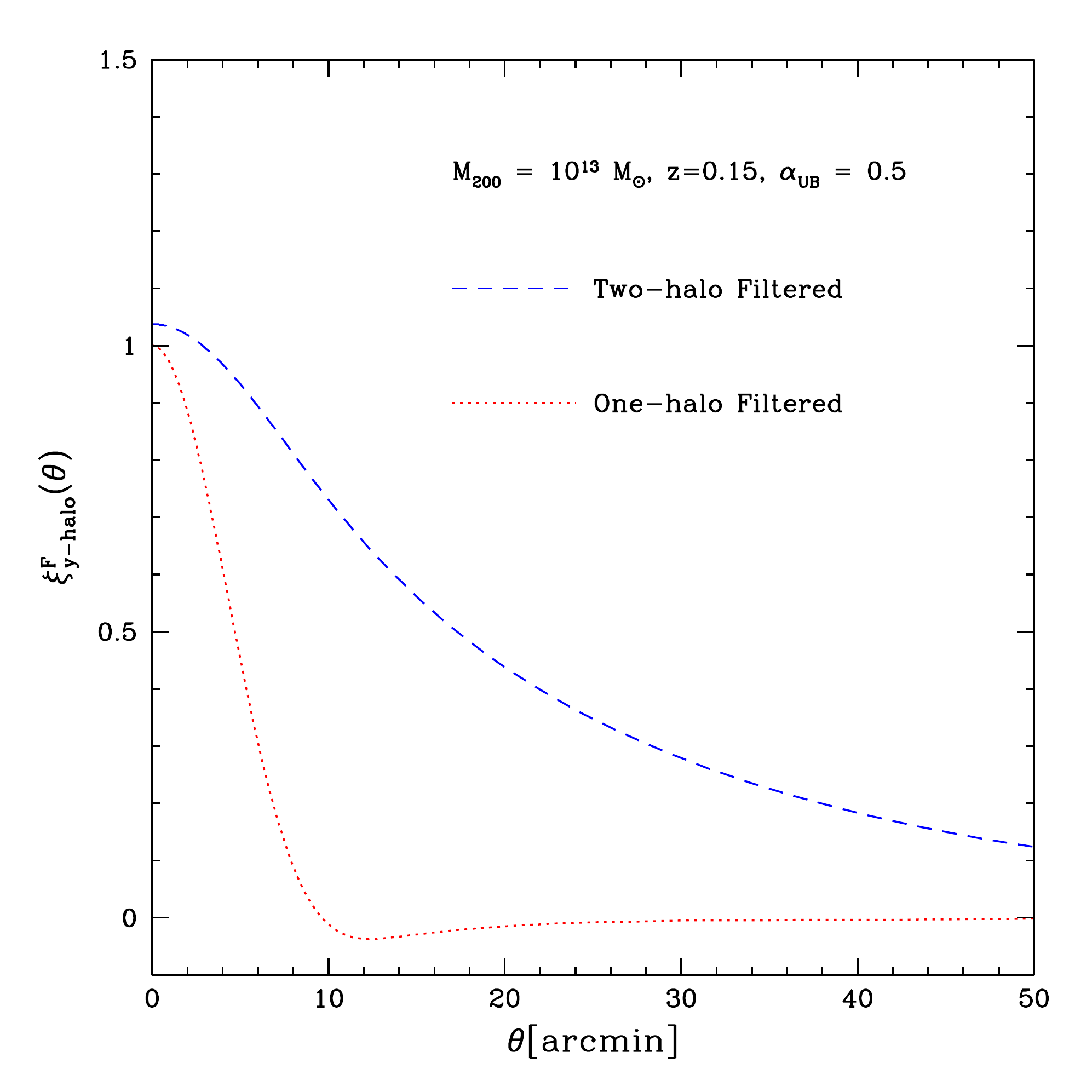}
\caption{The $y$-halo cross-correlation function for a halo of mass $M_{200} = 10^{13} M_\odot$ at $z=0.15$ for the {\it UB} model with $\alpha_{ub}=0.5$.  The left panel shows the one- (red dotted) and two-halo (blue dashed) terms before applying the matched filter, while the right panel shows the same quantities afterward.  In each case, the one-halo term is normalized to unity at $\theta \rightarrow 0$. The matched filter diminishes the impact of the two-halo term, but it still makes a strong contribution in this model.}
\label{fig:yprof_full}
\end{center}
\end{figure}

Fig.~\ref{fig:yprof_full} further shows the impact of the matched filter on the full $y$-halo cross-correlation functions for a halo of mass $M_{200}=10^{13} M_\odot$ at $z=0.15$ in the $\alpha_{ub}=0.5$ model. The matched filter suppresses the relative strength of the two-halo term, but in this model it is still dominant. Note also that the fractional importance of the two-halo term increases with $\theta$, so the diagnostic of Fig.~\ref{fig:ratio} is illustrative, but incomplete.

An alternate choice of filter could be used to better extract the one-halo term from the two-halo ``contamination'', at the expense of increased variance. For example, the two-halo term could be incorporated into the noise term in the denominator of Eq.~\ref{eq:match} to downweight modes where it is large relative to the one-halo contribution. Alternatively, $F({\sc l})$ could be multiplied by a simple high-pass filter. For example, multiplying $F({\sc l})$ by a high-pass filter that completely nulls modes with ${\sc l} \leq {\sc l_0}$ reduces the fractional two-halo contamination in the {\it UB} model with $\alpha_{ub} = 0.5$ (relative to the total cross-correlation function) at $\theta \rightarrow 0$ from $50\%$ to $31\%$ for ${\sc l_0}=300$, to $18\%$ for ${l_0}=600$, and to $10\%$ for ${l_0}=900$ (at the expense of increased variance).

Provided the filters adopted by~\citetalias{PlanckLBG2013} are similar to those shown in Fig.~\ref{fig:filter_examples}, it seems they still yield appreciable two-halo contributions, although this may be diminished by the isolation criteria used in defining the LBG sample.  For example, in the lower stellar mass bin considered in the left-hand panel of Fig.~\ref{fig:LoBaG}, the small-scale $y$-group cross-correlation function decreases by a factor of $\approx 2$ when the isolation radius is increased from $1$ to $2$ Mpc, and stabilizes at larger isolation radius. This suggests that the 1 Mpc-isolated sample receives comparable contributions from the one and two-halo terms at small radii. Note that this analysis is applied to the full Planck $y$-map. If the matched filter indeed suppresses the two-halo term by an {\em additional} factor of three, then the isolated, filtered two-halo contribution may be just $\approx 1/3$ of the one-halo term in the inner bins.  The fractional importance of the two-halo term increases with increasing radius/angle, but this may be partly compensated by the stronger suppression from the isolation. A $\approx 30\%$ bias may not be extreme cause for concern given the statistical errors in~\citetalias{PlanckLBG2013} at low stellar mass, but this is only a rough estimate.  Our conclusion is that two-halo contributions should be taken into account in future Compton-$y$ stacking or cross-correlation analyses, particularly those extending below cluster mass scales.
\end{appendix}


\end{document}